# Binary Hypothesis Testing via Measure Transformed Quasi Likelihood Ratio Test

Nir Halay*, Koby Todros* and Alfred O. Hero†
*Ben-Gurion University of the Negev, †University of Michigan

**Abstract**

In this paper, the Gaussian quasi likelihood ratio test (GQLRT) for non-Bayesian binary hypothesis testing is generalized by applying a transform to the probability distribution of the data. The proposed generalization, called measure-transformed GQLRT (MT-GQLRT), selects a Gaussian probability model that best empirically fits a transformed probability measure of the data. By judicious choice of the transform we show that, unlike the GQLRT, the proposed test is resilient to outliers and involves higher-order statistical moments leading to significant mitigation of the model mismatch effect on the decision performance. A Bayesian extension of the proposed MT-GQLRT is also developed that is based on selection of a Gaussian probability model that best empirically fits a transformed conditional probability distribution of the data. The non-Bayesian and Bayesian MT-GQLRTs are applied to signal detection and classification, in simulation examples that illustrate their advantages over the standard GQLRT and other robust alternatives.

## I. Introduction

Classical binary hypothesis testing deals with deciding between two hypotheses based on a sequence of multivariate samples from an underlying probability distribution that is equal to one of two known probability measures [1]. When the probability distributions under each hypothesis are correctly specified

This work was partially supported by United States-Israel Binational Science Foundation grant 2014334 and by ARO grant W911NF-15-1-0479.

the likelihood ratio test (LRT), which is the most powerful test for a given size[1] [2], can be implemented. In many practical scenarios the probability distributions are only partially known, and therefore, one must resort to suboptimal tests that utilize partial statistical information.

A popular suboptimal test of this kind is the Gaussian quasi LRT (GQLRT) [3]–[8] which assumes that the samples obey Gaussian distributions under each hypothesis. The GQLRT operates by selecting the Gaussian probability model that best empirically fits the data. When the observations are i.i.d. this selection is carried out by comparing the empirical Kullback-Leibler divergences [9] between the underlying probability distribution and the assumed normal probability measures. The GQLRT has gained popularity due to its implementation simplicity, ease of performance analysis, and its geometrical interpretations that arise from the convenient Gaussian model. Despite the model mismatch, introduced by the normality assumption, the GQLRT has the appealing property of consistency when the mean vectors and covariance matrices are correctly specified and identifiable over the considered hypotheses [6]. However, in some circumstances, such as for certain types of non-Gaussian data, large deviation from normality can inflict poor decision performance. This can occur when the first and second-order statistical moments are weakly identifiable over the considered hypotheses, or in the case of heavy-tailed data when the non-robust sample mean and covariance provide poor estimates in the presence of outliers.

To overcome these limitations, several alternatives have been proposed in the literature. One straightforward approach is a non-Gaussian quasi LRT (NGQLRT) that involves more complex distributional models, e.g., elliptical, at the possible expense of increased implementation complexity, cumbersome performance analysis, and degraded performance under nominal Gaussian data. For example, by assuming Laplace distributed observations the NGQLRT for weak DC signal detection in additive i.i.d. noise is the well established sign detector [10], [11]. Although the sign detector is more resilient against heavy-tailed noise outliers as compared to the GQLRT, it has considerably poor performance when the noise is Gaussian [10]. Another approach is based on higher-order cumulants [12], [13] that may improve identifiability. However, unlike the first and second-order cumulants, used in the GQLRT, these quantities involve complicated tensor analysis [14]. Additionally, their empirical estimates are highly non-robust to outliers and have increased computational and sample complexity.

In this paper, a robust generalization of the GQLRT is proposed that operates by selecting a Gaussian probability model that has the best empirical fit to a transformed probability distribution of the data. Under the proposed generalization, outlier-resilient tests can be obtained that involve higher-order statistical moments, and yet have the computational and implementation advantages of the standard GQLRT.

[1]In other words, the LRT attains the maximum detection probability (power) for a fixed false alarm rate (size).

This generalization, called the measure-transformed GQLRT (MT-GQLRT), is based on the measure transformation framework that was recently applied to canonical correlation analysis [15], [16], multiple signal classification (MUSIC) [17], [18] and parameter estimation [19], [20].

The considered measure transform is structured by a non-negative function, called the MT-function, and maps the probability distribution into a set of new probability measures on the observation space. By modifying the MT-function, classes of measure transformations can be obtained that have different useful properties that mitigate the model mismatch effect on the decision performance. Under the considered transform we redefine the measure-transformed (MT) mean vector and covariance matrix and show their relation to higher-order statistical moments. Furthermore, we reformulate the empirical estimates of the MT-mean and MT-covariance and restate the conditions on the MT-function for strong consistency and robustness to outliers. These quantities are then used to construct the proposed MT-GQLRT.

Similarly to the GQLRT, the proposed MT-GQLRT compares the empirical Kullback-Leibler divergences between probability distributions. The difference is that the MT-GQLRT compares the Kullback-Leibler divergences between the *transformed* probability distribution of the data and two normal probability measures that are characterized by the MT-mean vector and MT-covariance matrix under each hypothesis. Under some mild regularity conditions we show that the MT-GQLRT is consistent and its corresponding test statistic is asymptotically normal. Furthermore, given two training sequences from the probability distribution under each hypothesis, a data-driven procedure for optimal selection of the MT-function within some parametric class of functions is developed that maximizes an empirical estimate of the asymptotic power given a fixed empirical asymptotic size.

The proposed MT-GQLRT has the following properties that motivate its use: 1) Similarly to the standard GQLRT, the test-statistic of the proposed test has a simple closed-form expression that only involves mean vectors and covariance matrices. 2) For any non-constant analytic MT-function, the MT-mean vectors and MT-covariance matrices, comprising the test-statistic of the proposed MT-GQLRT, involve higher-order statistical moments. This can significantly improve the decision performance, comparing to the standard GQLRT, when the first and second-order statistical moments are weakly identifiable over the considered hypotheses. 3) Under some mild regularity conditions on the MT-function, we show that the empirical MT-mean and MT-covariance, comprising the test-statistic of the MT-GQLRT, are robust against outliers. This property can significantly improve the decision performance in the presence of heavy-tailed noise. 4) The performance analysis of the proposed test is tractable, which enables derivation of simple procedures for threshold determination and optimization of the MT-function parameters.

We go on to introduce a Bayesian extension of the proposed MT-GQLRT to mitigate the sensitivity of the standard Bayesian GQLRT [21]–[24] to model mismatch. Similarly to the non-Bayesian case, the Bayesian MT-GQLRT compares the empirical Kullback-Leibler divergences between a transformed conditional probability distribution of the data and two normal probability measures that are characterized by the MT-mean vector and MT-covariance matrix conditioned on each hypothesis. Like the non-Bayesian MT-GQLRT, the Bayesian MT-GQLRT can gain robustness against outliers under the same conditions on the MT-function and its corresponding test-statistic is asymptotically normal. Furthermore, given two training sequences from the conditional probability distribution of each hypothesis, optimal selection of a parametric MT-function and the threshold value is carried out via joint minimization of the empirical asymptotic Bayes risk [1].

The proposed MT-GQLRT and its Bayesian extension are illustrated for signal detection and classification, respectively, in the presence of spherically contoured noise. By specifying the MT-function within the family of zero-centered Gaussian functions parameterized by a scale parameter, we show that the MT-GQLRT can significantly mitigate the model mismatch effect introduced by the normality assumption. More specifically, we show that the proposed MT-GQLRT outperforms the non-robust GQLRT and other robust alternatives and attains decision performance that are significantly closer to those obtained by the omniscient LRT that, unlike the proposed test, requires complete knowledge of the likelihood functions under each hypothesis. In these application examples, we also provide suboptimal implementations of the non-Bayesian and Bayesian MT-GQLRT that do not require training sequences for selection of tuning parameters. We show that these training-sequence-free versions outperform the other robust tests considered that do not require training sequences.

The paper is organized as follows. In Section II, we formulate the considered hypothesis testing problem and review the GQLRT. Section III reviews the principles of the considered probability measure transform. In Section IV, we use this transformation to construct the non-Bayesian MT-GQLRT. The extension for Bayesian hypothesis testing is developed in section V. The MT-GQLRT and its Bayesian extension are applied to signal detection and classification, respectively, in Section VI. In Section VII, the main points of this contribution are summarized. The proofs of the propositions and theorems stated throughout the paper are given in the Appendix.

## II. Preliminaries

In this section, we formulate the considered binary hypothesis testing problem. We proceed by reviewing the GQLRT [3]–[8]. We show that the GQLRT can be interpreted as a comparison between the empirical

Kullback-Leibler divergences between the probability distribution of the data and two normal probability measures. This operation principle will be used in Section IV to develop the proposed MT-GQLRT.

### A. Problem formulation

We define the measure space $(\mathcal{X}, \mathcal{S}, P)$, where $\mathcal{X} \subseteq \mathbb{C}^p$ is the observation space of a complex-valued random vector $\mathbf{X}$, $\mathcal{S}$ is a $\sigma$-algebra over $\mathcal{X}$ and $P$ is a probability measure on $\mathcal{S}$ which belongs to a pair set $\{P_0, P_1\}$. It is assumed that $P$ is absolutely continuous w.r.t. a dominating $\sigma$-finite measure $\rho$ on $\mathcal{S}$, such that the Radon-Nikodym derivatives [25]

$$f(\mathbf{x}) \triangleq \frac{dP(\mathbf{x})}{d\rho(\mathbf{x})} \tag{1}$$

exists. The function $f(\cdot)$ is called the density function of $P$. Let $g : \mathcal{X} \rightarrow \mathbb{C}$ denote an integrable scalar function. The expectation of $g(\mathbf{X})$ under $P$ is defined as:

$$\mathrm{E}[g(\mathbf{X}); P] \triangleq \int_{\mathcal{X}} g(\mathbf{x})\, dP(\mathbf{x}),$$

where $\mathbf{x} \in \mathcal{X}$.

Given a sequence of samples $\mathbf{X}_n, n = 1, ..., N$ from $P$ we consider the problem of testing between the null and alternative hypotheses

$$\begin{aligned} H_0 &: P = P_0 \\ H_1 &: P = P_1, \end{aligned} \tag{2}$$

respectively, when $P_0$ and $P_1$ are partially known. The GQLRT, that is reviewed in the following subsection, assumes that partial statistical information is available through the standard mean vectors and the covariance matrices under each hypothesis. The proposed MT-GQLRT that will be developed in Section IV exploits higher-order moment information through measure-transformed mean vectors and covariance matrices.

### B. Review of the Gaussian quasi likelihood ratio test

Let $\Phi_k$, $k = 0, 1$ denote two proper complex Gaussian probability measures that are characterized by the mean vectors $\boldsymbol{\mu}_k \triangleq \mathrm{E}[\mathbf{X}; P_k]$, $k = 0, 1$, and the covariance matrices $\boldsymbol{\Sigma}_k \triangleq \mathrm{E}\left[\mathbf{X}\mathbf{X}^H; P_k\right] - \boldsymbol{\mu}_k \boldsymbol{\mu}_k^H$, $k = 0, 1$. Given a sequence of samples $\mathbf{X}_n, n = 1, ..., N$ from the underlying probability distribution $P$, the GQLRT applies LRT under the assumption that $P_0 = \Phi_0$ and $P_1 = \Phi_0$, which leads to the following decision rule:

$$
\begin{aligned}
T &\triangleq \frac{1}{N}\sum_{n=1}^{N} \psi\left(\mathbf{X}_n\right) \quad (3)\\
&= \left(D_{\rm LD}\left[\hat{\mathbf{\Sigma}}||\mathbf{\Sigma}_0\right] + \|\hat{\boldsymbol{\mu}} - \boldsymbol{\mu}_0\|^2_{\mathbf{\Sigma}_0^{-1}}\right) - \left(D_{\rm LD}\left[\hat{\mathbf{\Sigma}}||\mathbf{\Sigma}_1\right] + \|\hat{\boldsymbol{\mu}} - \boldsymbol{\mu}_1\|^2_{\mathbf{\Sigma}_1^{-1}}\right) \underset{H_0}{\overset{H_1}{\gtrless}} t,
\end{aligned}
$$

where $\psi(\mathbf{X}) \triangleq \log(\phi_1(\mathbf{X})/\phi_0(\mathbf{X}))$, and

$$
\phi_k\left(\mathbf{x}\right) \triangleq \exp\left(-\left(\mathbf{x} - \boldsymbol{\mu}_k\right)^H \mathbf{\Sigma}_k^{-1}\left(\mathbf{x} - \boldsymbol{\mu}_k\right)\right) / \det\left[\pi\mathbf{\Sigma}_k\right] \quad (4)
$$

is the density function of $\Phi_k$ w.r.t. the dominating $\sigma$-finite measure $\rho$ on $\mathcal{S}$. In the second equality of (3) $D_{\rm LD}[\mathbf{A}||\mathbf{B}] \triangleq {\rm tr}[\mathbf{A}\mathbf{B}^{-1}] - \log\det[\mathbf{A}\mathbf{B}^{-1}] - p$ is the log-determinant divergence [26] between positive definite matrices $\mathbf{A}, \mathbf{B}$, $\|\mathbf{a}\|_{\mathbf{C}} \triangleq \sqrt{\mathbf{a}^H\mathbf{C}\mathbf{a}}$ denotes the weighted Euclidean norm of a vector $\mathbf{a}$ with positive-definite weighting matrix $\mathbf{C}$ and $\hat{\boldsymbol{\mu}} \triangleq \frac{1}{N}\sum_{n=1}^{N}\mathbf{X}_n$ and $\hat{\mathbf{\Sigma}} \triangleq \frac{1}{N}\sum_{n=1}^{N}\mathbf{X}_n\mathbf{X}_n^H - \hat{\boldsymbol{\mu}}\hat{\boldsymbol{\mu}}^H$ denote the standard sample mean vector (SMV) and sample covariance matrix (SCM). The parameter $t \in \mathbb{R}$ denotes a threshold.

In the following we show that the GQLRT (3) operates by comparing the empirical Kullback-Leibler divergences between $P$ and $\Phi_k$, $k = 0, 1$. The Kullback-Leibler divergence (KLD) between $P$ and $\Phi_k$, $k \in \{0, 1\}$ is defined as [9]:

$$
D_{\rm KL}\left[P||\Phi_k\right] \triangleq {\rm E}\left[\log\frac{f\left(\mathbf{X}\right)}{\phi_k\left(\mathbf{X}\right)}; P\right], \quad (5)
$$

where $f(\cdot)$ is the density function (1) of $P$. Given a sequence of samples $\mathbf{X}_n, n = 1, ..., N$ from $P$, an empirical estimate of (5) is defined as:

$$
\hat{D}_{\rm KL}\left[P||\Phi_k\right] \triangleq \frac{1}{N}\sum_{n=1}^{N}\log\frac{f\left(\mathbf{X}_n\right)}{\phi_k\left(\mathbf{X}_n\right)}.
$$

Hence, the difference $\hat{D}_{\rm KL}[P||\Phi_0] - \hat{D}_{\rm KL}[P||\Phi_1]$ coincides with the test statistic in (3). Finally, by (4) and (5), one can verify that when $\phi_0(\cdot) \neq \phi_1(\cdot)$, the difference $D_{\rm KL}[P||\Phi_0] - D_{\rm KL}[P||\Phi_1]$ will be negative if $P = P_0$ and positive if $P = P_1$. This information-theoretic interpretation provides justification for the test statistic (3).

## III. Probability measure transform: Review

In this section, we review the principles of the probability measure transform [15]–[20]. We redefine the measure-transformed mean vector and covariance matrix and show their relation to higher-order statistical moments. Moreover, we reformulate their empirical estimators and restate the conditions for strong consistency and robustness to outliers. These quantities will be used in the following section to construct the measure-transformed GQLRT.

### A. Probability measure transform

**Definition 1.** *Given a non-negative function* $u : \mathbb{C}^p \to \mathbb{R}_+$ *satisfying*

$$0 < \mathrm{E}\left[u\left(\mathbf{X}\right); P\right] < \infty, \tag{6}$$

*a transform on* $P$ *is defined via the relation:*

$$Q^{(u)}\left(A\right) \triangleq \mathbb{T}_u\left[P\right]\left(A\right) \triangleq \int_A \varphi_u\left(\mathbf{x}\right) dP\left(\mathbf{x}\right), \tag{7}$$

*where* $A \in \mathcal{S}$ *and*

$$\varphi_u\left(\mathbf{x}\right) \triangleq \frac{u\left(\mathbf{x}\right)}{\mathrm{E}\left[u\left(\mathbf{X}\right); P\right]}. \tag{8}$$

*The function* $u\left(\cdot\right)$ *is called the MT-function.*

By definition 1, one can verify that $Q^{(u)}$ is a probability measure on $\mathcal{S}$ that is absolutely continuous w.r.t. $P$, with Radon-Nikodym derivative [25]:

$$\frac{dQ^{(u)}\left(\mathbf{x}\right)}{dP\left(\mathbf{x}\right)} = \varphi_u\left(\mathbf{x}\right). \tag{9}$$

The MT-function $u(\cdot)$ is the generating function of the probability measure $Q^{(u)}$. By modifying $u(\cdot)$ a wide range of probability measures on $\mathcal{S}$ can be obtained.

### B. The MT-mean and MT-covariance

According to (9) the mean vector and covariance matrix of $\mathbf{X}$ under $Q^{(u)}$ are given by:

$$\boldsymbol{\mu}^{(u)} \triangleq \mathrm{E}[\mathbf{X}; Q^{(u)}] = \mathrm{E}\left[\mathbf{X}\varphi_u\left(\mathbf{X}\right); P\right] \tag{10}$$

and

$$\boldsymbol{\Sigma}^{(u)} \triangleq \mathrm{cov}[\mathbf{X}; Q^{(u)}] = \mathrm{E}\left[\mathbf{X}\mathbf{X}^H\varphi_u\left(\mathbf{X}\right); P\right] - \boldsymbol{\mu}^{(u)}\boldsymbol{\mu}^{(u)H}, \tag{11}$$

respectively. Equations (10) and (11) imply that $\boldsymbol{\mu}^{(u)}$ and $\boldsymbol{\Sigma}^{(u)}$ are weighted mean and covariance of $\mathbf{X}$ under $P$, with the weighting function $\varphi_u(\cdot)$ defined in (8). Notice that when the MT-function $u(\cdot)$ is non-zero and constant valued, the standard mean vector $\boldsymbol{\mu}$ and covariance matrix $\boldsymbol{\Sigma}$ are obtained. Alternatively, when $u(\cdot)$ is a non-constant analytic function, which has a convergent Taylor series expansion, the resulting MT-mean and MT-covariance involve *higher-order statistical moments* of $P$.

### C. The empirical MT-mean and MT-covariance

Given a sequence of $N$ i.i.d. samples from $P$, the empirical estimators of $\boldsymbol{\mu}^{(u)}$ and $\boldsymbol{\Sigma}^{(u)}$ are defined as:

$$\hat{\boldsymbol{\mu}}^{(u)} \triangleq \sum_{n=1}^{N} \mathbf{X}_n \hat{\varphi}_u\left(\mathbf{X}_n\right) \tag{12}$$

and

$$\hat{\boldsymbol{\Sigma}}^{(u)} \triangleq \sum_{n=1}^{N} \mathbf{X}_n \mathbf{X}_n^H \hat{\varphi}_u\left(\mathbf{X}_n\right) - \hat{\boldsymbol{\mu}}^{(u)} \hat{\boldsymbol{\mu}}^{(u)H}, \tag{13}$$

respectively, where

$$\hat{\varphi}_u\left(\mathbf{X}_n\right) \triangleq \frac{u\left(\mathbf{X}_n\right)}{\sum_{j=1}^{N} u\left(\mathbf{X}_j\right)}. \tag{14}$$

According to Proposition 2 in [17], if $\mathrm{E}[\|\mathbf{X}\|^2 u(\mathbf{X}); P] < \infty$ then $\hat{\boldsymbol{\mu}}^{(u)} \xrightarrow{\text{w.p.1}} \boldsymbol{\mu}^{(u)}$ and $\hat{\boldsymbol{\Sigma}}^{(u)} \xrightarrow{\text{w.p.1}} \boldsymbol{\Sigma}^{(u)}$ as $N \to \infty$, where "$\xrightarrow{\text{w.p.1}}$" denotes convergence with probability (w.p.) 1 [27]. Note that for $u\left(\mathbf{X}\right) \equiv 1$ the estimators $\hat{\boldsymbol{\mu}}^{(u)}$ and $\frac{N}{N-1}\hat{\boldsymbol{\Sigma}}^{(u)}$ reduce to the standard unbiased sample mean vector (SMV) and sample covariance matrix (SCM), respectively. Finally, we note that by [28]–[30] it follows that (12) and (13) are different than M-estimators of location and scatter that use different weight functions and generally implemented as an iterative fixed-point algorithm.

### D. Robustness to outliers

Robustness of the empirical MT-covariance (13) to outliers was studied in [17] using its influence function [31], which describes the bias effect on the estimator introduced by an infinitesimal contamination at some point $\mathbf{y} \in \mathbb{C}^p$. An estimator is said to be B-robust if its influence function is bounded [31]. Similarly to the proof of Proposition 3 in [17] it can be shown that if there exists a finite positive constant $M \in \mathbb{R}$, such that for all $\mathbf{y} \in \mathbb{C}^p$:

$$u(\mathbf{y}) \leq M \qquad \text{and} \qquad u(\mathbf{y})\|\mathbf{y}\|^2 \leq M, \tag{15}$$

then the influence functions of both (12) and (13) are bounded.

## IV. The Measure-Transformed Gaussian quasi likelihood ratio test

In this section, we extend the GQLRT (3) by applying the transformation (7) to the underlying probability measure $P$. Here, we assume that partial statistical information is available through the MT-mean vectors and the MT-covariance matrices under each hypothesis. Regularity conditions for asymptotic normality of the proposed test statistic are derived. When these conditions are satisfied we show that the resulting test is consistent and derive its asymptotic size and power. Optimal selection of the MT-function $u(\cdot)$ out of some parametric class of functions is also discussed. Finally, we describe the steps for implementation of the proposed MT-GQLRT.

### A. The MT-GQLRT

Similarly to the standard GQLRT, given a sequence of samples from $P$, the proposed MT-GQLRT compares the empirical KLDs between the transformed probability distribution of the data $Q^{(u)}$ (which can be either $Q_0^{(u)} \triangleq \mathbb{T}_u[P_0]$ if $H_0$ is true or $Q_1^{(u)} \triangleq \mathbb{T}_u[P_1]$ if $H_1$ is true) and two proper complex Gaussian probability measures $\Phi_k^{(u)}$, $k = 0, 1$ that are characterized by the MT-mean vectors $\boldsymbol{\mu}_k^{(u)} \triangleq \mathrm{E}[\mathbf{X}; Q_k^{(u)}]$, $k = 0, 1$ and the MT-covariance matrices $\boldsymbol{\Sigma}_k^{(u)} \triangleq \mathrm{cov}[\mathbf{X}; Q_k^{(u)}]$, $k = 0, 1$. The KLD between $Q^{(u)}$ and $\Phi_k^{(u)}$, $k \in \{0, 1\}$ is defined as [9]:

$$D_{\mathrm{KL}}\left[Q^{(u)}||\Phi_k^{(u)}\right] \triangleq \mathrm{E}\left[\log \frac{q^{(u)}(\mathbf{X})}{\phi_k^{(u)}(\mathbf{X})}; Q^{(u)}\right], \tag{16}$$

where $q^{(u)}(\cdot)$ is the unknown density of $Q^{(u)}$ w.r.t. the dominating measure $\rho$ on $\mathcal{S}$, and

$$\phi_k^{(u)}(\mathbf{x}) \triangleq \exp\left(-(\mathbf{x} - \boldsymbol{\mu}_k^{(u)})^H (\boldsymbol{\Sigma}_k^{(u)})^{-1} (\mathbf{x} - \boldsymbol{\mu}_k^{(u)})\right) / \det\left[\pi \boldsymbol{\Sigma}_k^{(u)}\right] \tag{17}$$

is the density of $\Phi_k^{(u)}$ w.r.t. $\rho$. By (16) and (17), one can verify that when $\phi_0^{(u)}(\cdot) \neq \phi_1^{(u)}(\cdot)$, the difference $D_{\mathrm{KL}}[Q^{(u)}||\Phi_0^{(u)}] - D_{\mathrm{KL}}[Q^{(u)}||\Phi_1^{(u)}]$ will be negative under $H_0$ and positive under $H_1$. Hence, similarly to the standard GQLRT, this justifies the use of the empirical estimate of this difference as a test statistic for testing $H_0$ versus $H_1$.

According to (9), the divergence $D_{\mathrm{KL}}[Q^{(u)}||\Phi_k^{(u)}]$, $k \in \{0, 1\}$ can be estimated using only samples from $P$. Therefore, similarly to (12) and (13), an empirical estimate of (16) given a sequence of samples $\mathbf{X}_n, n = 1, ..., N$ from $P$, is defined as:

$$\hat{D}_{\mathrm{KL}}\left[Q^{(u)}||\Phi_k^{(u)}\right] \triangleq \sum_{n=1}^{N} \hat{\varphi}_u(\mathbf{X}_n) \log \frac{q^{(u)}(\mathbf{X}_n)}{\phi_k^{(u)}(\mathbf{X}_n)},$$

where $\hat{\varphi}_u(\cdot)$ is defined in (14). Thus, the proposed test statistic, which is independent of the unknown transformed density function $q^{(u)}(\mathbf{x})$, is defined as:

$$\begin{aligned} T_u &\triangleq \hat{D}_{\mathrm{KL}}[Q^{(u)}||\Phi_0^{(u)}] - \hat{D}_{\mathrm{KL}}[Q^{(u)}||\Phi_1^{(u)}] = \sum_{n=1}^{N} \hat{\varphi}_u(\mathbf{X}_n)\psi_u(\mathbf{X}_n) \\ &= \left(D_{\mathrm{LD}}\left[\hat{\boldsymbol{\Sigma}}^{(u)}||\boldsymbol{\Sigma}_0^{(u)}\right] + \left\|\hat{\boldsymbol{\mu}}^{(u)} - \boldsymbol{\mu}_0^{(u)}\right\|^2_{(\boldsymbol{\Sigma}_0^{(u)})^{-1}}\right) \\ &- \left(D_{\mathrm{LD}}\left[\hat{\boldsymbol{\Sigma}}^{(u)}||\boldsymbol{\Sigma}_1^{(u)}\right] + \left\|\hat{\boldsymbol{\mu}}^{(u)} - \boldsymbol{\mu}_1^{(u)}\right\|^2_{(\boldsymbol{\Sigma}_1^{(u)})^{-1}}\right), \end{aligned} \tag{18}$$

where

$$\psi_u(\mathbf{X}) \triangleq \log\left(\phi_1^{(u)}(\mathbf{X})/\phi_0^{(u)}(\mathbf{X})\right), \tag{19}$$

and the operators $D_{\rm LD}[\cdot||\cdot]$ and $\|\cdot\|_{(\cdot)}$ are defined below (4). The decision rule based on the test statistic (18) is:

$$T_u \underset{H_0}{\overset{H_1}{\gtrless}} t, \tag{20}$$

where $t \in \mathbb{R}$ denotes a threshold. By modifying the MT-function $u(\cdot)$ such that condition (6) is satisfied the MT-GQLRT is modified, resulting in a family of tests generalizing the GQLRT (3). In particular, if $u(\cdot)$ is any non-zero constant function over $\mathcal{X}$, then $Q^{(u)} = P$ and the standard non-robust GQLRT is obtained which only involves first and second-order statistical moments. Otherwise, when $u(\cdot)$ is a non-constant analytic function that satisfies condition (15), the resulting test is outlier resilient and involves higher-order statistical moments.

### B. Asymptotic performance analysis

Here, we study the asymptotic decision performance of the proposed MT-GQLRT (20). For simplicity, we assume that a sequence of i.i.d. samples $\mathbf{X}_n$, $n = 1, \ldots, N$ from $P$ is available.

**Theorem 1** (Asymptotic normality)**.** *Assume that the following conditions are satisfied:*

*A-1)* $\boldsymbol{\mu}_0^{(u)} \neq \boldsymbol{\mu}_1^{(u)}$ *or* $\boldsymbol{\Sigma}_0^{(u)} \neq \boldsymbol{\Sigma}_1^{(u)}$.

*A-2)* $\boldsymbol{\Sigma}_0^{(u)}$ *and* $\boldsymbol{\Sigma}_1^{(u)}$ *are non-singular.*

*A-3)* $\mathrm{E}[u^2(\mathbf{X}); P]$ *and* $\mathrm{E}[\|\mathbf{X}\|^4 u^2(\mathbf{X}); P]$ *are finite for* $P = P_0$ *and* $P = P_1$.

*Then,*

$$\frac{T_u - \eta_k^{(u)}}{\sqrt{\lambda_k^{(u)}}} \xrightarrow[N\to\infty]{D} \mathcal{N}(0,1),$$

*where "$\xrightarrow{D}$" denotes convergence in distribution [27],*

$$\eta_k^{(u)} \triangleq \mathrm{E}\left[\varphi_{u,k}(\mathbf{X})\,\psi_u(\mathbf{X}); P_k\right], \tag{21}$$

$$\lambda_k^{(u)} \triangleq \frac{1}{N}\mathrm{E}\left[\varphi_{u,k}^2(\mathbf{X})\left(\psi_u(\mathbf{X}) - \eta_k^{(u)}\right)^2; P_k\right]. \tag{22}$$

*and* $\varphi_{u,k}(\cdot)$ *is defined according to (8) with* $P$ *replaced by* $P_k$. *[A proof is given in Appendix B]*

**Corollary 1** (Asymptotic size and power)**.** *Assume that the conditions stated in Theorem 1 are satisfied. The asymptotic size and power of the decision rule (20) are given by:*

$$\alpha_u \triangleq Q\left(\frac{t - \eta_0^{(u)}}{\sqrt{\lambda_0^{(u)}}}\right) \quad \text{and} \quad \beta_u \triangleq Q\left(\frac{t - \eta_1^{(u)}}{\sqrt{\lambda_1^{(u)}}}\right), \tag{23}$$

*respectively, where $Q(\cdot)$ denotes the tail probability of the standard normal distribution [32].*

**Corollary 2** (Consistency)**.** *Assume that the conditions in Theorem 1 are satisfied. Then, for any fixed asymptotic size the asymptotic power of the test (20) satisfies $\beta_u \to 1$ as $N \to \infty$.*

In the following Proposition, strongly consistent estimates of the asymptotic size and power (23) are constructed based on two i.i.d. training sequences from $P_0$ and $P_1$. These will be used in the sequel for optimal selection of the MT-function.

**Proposition 1** (Empirical asymptotic size and power)**.** *Let $\mathbf{X}_n^{(k)}$, $n = 1, \ldots, N_k$, $k = 0, 1$ denote sequences of i.i.d. samples from $P_0$ and $P_1$, respectively. Define the empirical asymptotic size and power:*

$$\hat{\alpha}_u \triangleq Q\left(\frac{t - \hat{\eta}_0^{(u)}}{\sqrt{\hat{\lambda}_0^{(u)}}}\right) \quad \textit{and} \quad \hat{\beta}_u \triangleq Q\left(\frac{t - \hat{\eta}_1^{(u)}}{\sqrt{\hat{\lambda}_1^{(u)}}}\right), \tag{24}$$

*respectively, where*

$$\hat{\eta}_k^{(u)} \triangleq \sum_{n=1}^{N_k} \hat{\varphi}_u\left(\mathbf{X}_n^{(k)}\right) \psi_u\left(\mathbf{X}_n^{(k)}\right) \tag{25}$$

*and*

$$\hat{\lambda}_k^{(u)} \triangleq \frac{N_k}{N} \sum_{n=1}^{N_k} \hat{\varphi}_u^2\left(\mathbf{X}_n^{(k)}\right) \left(\psi_u\left(\mathbf{X}_n^{(k)}\right) - \hat{\eta}_k^{(u)}\right)^2. \tag{26}$$

*Assume that conditions A-1—A-3 stated in Theorem 1 are satisfied. Then,*

$$\hat{\alpha}_u \xrightarrow[N_0 \to \infty]{\text{w.p.1}} \alpha_u \quad \textit{and} \quad \hat{\beta}_u \xrightarrow[N_1 \to \infty]{\text{w.p.1}} \beta_u.$$

*[A proof is given in Appendix C]*

### *C. Selection of the MT-function*

When the observations are normally distributed the GQLRT (3) coincides with the LRT, which is the most powerful test for a fixed size (false alarm rate). Hence, in this case, following the discussion below Eq. (20), the optimal MT-function $u(\cdot)$ should be non-zero and constant valued. Unfortunately, in the non-Gaussian case finding the optimal MT-function associated with the asymptotically most powerful test for a fixed false alarm rate is analytically cumbersome and requires the knowledge of the likelihood functions under each hypothesis. Therefore, we propose to specify the MT-function within some parametric family $\{u(\mathbf{X}; \boldsymbol{\omega}), \boldsymbol{\omega} \in \boldsymbol{\Omega} \subseteq \mathbb{C}^r\}$ that satisfies the conditions stated in Definition 1 and Theorem 1. For example, in order to gain resilience against outliers, the Gaussian family of functions that satisfy condition (15) is a natural choice. An optimal choice of the MT-function parameter $\boldsymbol{\omega}$ would be the one that maximizes the empirical asymptotic power in (24) at a fixed empirical asymptotic size $\hat{\alpha}_u = \alpha$, i.e., we maximize the following objective function:

$$\hat{\beta}_u^{(\alpha)}(\boldsymbol{\omega}) = Q\left(\frac{\hat{\eta}_0^{(u)}(\boldsymbol{\omega}) - \hat{\eta}_1^{(u)}(\boldsymbol{\omega}) + \sqrt{\hat{\lambda}_0^{(u)}(\boldsymbol{\omega})}Q^{-1}(\alpha)}{\sqrt{\hat{\lambda}_1^{(u)}(\boldsymbol{\omega})}}\right). \quad (27)$$

Notice that in practice, it is sufficient to minimize the argument of $Q(\cdot)$ as it is monotonically decreasing.

### D. *Implementation of the test*

Given a sequence of data samples $\mathbf{X}_n$, $n = 1, \ldots, N$, two training sequences $\mathbf{X}_n^{(k)}$, $n = 1, \ldots, N_k$, $k = 0, 1$ from $P_0$ and $P_1$, and a class of MT-functions $\{u(\mathbf{X}; \boldsymbol{\omega}), \boldsymbol{\omega} \in \boldsymbol{\Omega} \subseteq \mathbb{C}^r\}$, the proposed MT-GQLRT is implemented via the following steps:

1) Fix an empirical asymptotic size $\hat{\alpha}_u = \alpha$.
2) Obtain the optimal MT-function parameter $\boldsymbol{\omega}_{\text{opt}}$ by maximizing (27) w.r.t. $\boldsymbol{\omega}$.
3) Compute the threshold using the formula $t = \hat{\eta}_0^{(u)}(\boldsymbol{\omega}_{\text{opt}}) + \sqrt{\hat{\lambda}_0^{(u)}(\boldsymbol{\omega}_{\text{opt}})}Q^{-1}(\alpha)$ that follows directly from (24).
4) Apply the decision rule (20).

The maximization in step (2) is carried out numerically. Note that for each candidate of $\boldsymbol{\omega}$ only four scalars need to be computed using (25) and (26) in order to obtain the objective function (27). Hence, when $\boldsymbol{\omega}$ is one-dimensional, a simple line search can be implemented. Otherwise, the maximization can be performed via gradient ascend [33] or via greedy search. These techniques are more computationally efficient than a direct multidimensional search. However, they do not guarantee convergence to a global maximum when the objective function (27) is multimodal.

## V. Bayesian Extension

In this section, we develop a Bayesian extension of the proposed MT-GQLRT (20). In difference to the non-Bayesian formulation (2), here, we assume that each hypothesis $H_k$, $k \in \{0, 1\}$ has a known prior probability $\pi_k$ and that $P_k$ is a conditional probability distribution of $\mathbf{X}$ given that $H_k$ is true. Here, the measure transformation $\mathbb{T}_u[\cdot]$ (7) is applied to the conditional distribution $P \in \{P_0, P_1\}$. In this context, it is important to note that the measure-transformation properties stated in Section III, for the unconditional distribution, also apply here, for the conditional distribution. Similarly to the non-Bayesian MT-GQLRT, developed in Subsection IV-A, given a sequence of samples $\mathbf{X}_n$, $n = 1, \ldots, N$ from the conditional distribution $P$, the Bayesian MT-GQLRT compares the empirical KLDs between the transformed conditional probability distribution $Q^{(u)} \triangleq \mathbb{T}_u[P]$ and two normal probability measures $\Phi_k$, $k = 0, 1$ that are characterized by the MT-mean vectors $\boldsymbol{\mu}_k^{(u)} \triangleq \mathrm{E}[\mathbf{X}; Q_k^{(u)}]$, $k = 0, 1$, and the MT-covariance matrices $\boldsymbol{\Sigma}_k^{(u)} \triangleq \mathrm{cov}[\mathbf{X}; Q_k^{(u)}]$, $k = 0, 1$, conditioned on $H_k$, $k = 0, 1$. Thus, one can easily verify that the decision rule of the Bayesian MT-GQLRT is the same as the one of the non-Bayesian MT-GQLRT (18). The difference is in the performance analysis that is quantified through the Bayes-Risk.

### *A. Asymptotic performance analysis*

As in the non-Bayesian case, we assume that a sequence of i.i.d. samples $\mathbf{X}_n$, $n = 1, \ldots, N$ from the conditional distribution $P \in \{P_0, P_1\}$ is available. Straight forward extension of Theorem 1 (asymptotic normality of the test statistic) to the considered Bayesian case can be obtained here. Under this extension, the asymptotic Bayes risk and its empirical estimate are stated in the following propositions.

**Proposition 2** (Asymptotic Bayes risk)**.** *Assume that under the conditional distributions $P_0$ and $P_1$, the assumptions in Theorem 1 are satisfied. Let $L_{jk}$, $j, k \in \{0, 1\}$, denote the loss for deciding $H_j$ when $H_k$ is true, where $L_{00} = L_{11} = 0$. The asymptotic Bayes risk can be written as:*

$$R^{(u)}(t) \triangleq L_{10}\pi_0 Q\left(\frac{t - \eta_0^{(u)}}{\sqrt{\lambda_0^{(u)}}}\right) + L_{01}\pi_1 Q\left(\frac{\eta_1^{(u)} - t}{\sqrt{\lambda_1^{(u)}}}\right), \tag{28}$$

*where $\eta_k^{(u)}$ and $\lambda_k^{(u)}$ are defined as in (21) and (22), respectively.*

In the following Proposition, a strongly consistent estimate of the asymptotic Bayes risk (28) is constructed based on two i.i.d. sequences from the conditional distributions $P_0$ and $P_1$. This quantity will be used in the sequel for optimal selection of the MT-function.

**Proposition 3** (Empirical asymptotic Bayes risk)**.** *Let $\mathbf{X}_n^{(k)}$, $n = 1, \ldots, N_k$, $k = 0, 1$ denote sequences of i.i.d. samples from the conditional distributions $P_0$ and $P_1$, respectively. Define the empirical asymptotic Bayes risk:*

$$\hat{R}^{(u)}(t) \triangleq L_{10}\pi_0 Q\left(\frac{t - \hat{\eta}_0^{(u)}}{\sqrt{\hat{\lambda}_0^{(u)}}}\right) + L_{01}\pi_1 Q\left(\frac{\hat{\eta}_1^{(u)} - t}{\sqrt{\hat{\lambda}_1^{(u)}}}\right), \tag{29}$$

*where $\hat{\eta}_k^{(u)}$ and $\hat{\lambda}_k^{(u)}$ are defined as in (25) and (26), respectively. Assume that conditions A-1–A-3 stated in Theorem 1 are satisfied for the conditional distributions $P_0$ and $P_1$. Then, $\hat{R}^{(u)} \xrightarrow{w.p.1} R^{(u)}$ as $N_0, N_1 \to \infty$. [A proof is given in Appendix D]*

In the following Proposition, a necessary and sufficient condition for existence and uniqueness of an optimal threshold minimizing the empirical asymptotic Bayes risk (29) is derived. A closed form expression of this threshold is also presented.

**Proposition 4** (Optimal threshold). *Assume that* $\hat{\lambda}_0^{(u)} \neq \hat{\lambda}_1^{(u)}$[2]. *Define*

$$\hat{s}^{(u)} \triangleq \left(\hat{\eta}_0^{(u)} - \hat{\eta}_1^{(u)}\right)^2 - 2\left(\hat{\lambda}_0^{(u)} - \hat{\lambda}_1^{(u)}\right) \log \frac{L_{10}\pi_0\sqrt{\hat{\lambda}_1^{(u)}}}{L_{01}\pi_1\sqrt{\hat{\lambda}_0^{(u)}}}.$$

*A global minimum of the empirical asymptotic Bayes risk (29) exists and given by*

$$t_{opt}^{(u)} \triangleq \frac{\hat{\lambda}_0^{(u)}\hat{\eta}_1^{(u)} - \hat{\lambda}_1^{(u)}\hat{\eta}_0^{(u)} - \sqrt{\hat{\lambda}_0^{(u)}\hat{\lambda}_1^{(u)}\hat{s}^{(u)}}}{\hat{\lambda}_0^{(u)} - \hat{\lambda}_1^{(u)}} \tag{30}$$

*if and only if C-1)* $\hat{s}^{(u)} \geq 0$ *and C-2) the empirical Bayes risk (29) satisfies* $\hat{R}^{(u)}(t_{opt}^{(u)}) < \min\left(L_{10}\pi_0, L_{01}\pi_1\right)$. *[A proof is given in Appendix E]*

### B. Optimal selection of the MT-function

Similarly to the non-Bayesian case, we propose to specify the MT-function within some parametric family $\{u(\mathbf{X}; \boldsymbol{\omega}), \boldsymbol{\omega} \in \boldsymbol{\Omega} \subseteq \mathbb{C}^r\}$ of functions that have strictly positive and finite expectation w.r.t. the conditional distribution of the data and satisfy the conditions stated in Proposition 2. An optimal choice of the MT-function parameter $\boldsymbol{\omega}$ minimizes the empirical asymptotic Bayes risk (29) evaluated at the optimal threshold (30).

### C. Implementation of the test

Given the a-priori probabilities $\pi_0$ and $\pi_1$, the loss coefficients $L_{01}$ and $L_{10}$, a sequence of data samples $\mathbf{X}_n$, $n = 1, \ldots, N$ from $P$, two training sequences $\mathbf{X}_n^{(k)}$, $n = 1, \ldots, N_k$, $k = 0, 1$ from $P_0$ and $P_1$, and a class of MT-functions $\{u(\mathbf{X}; \boldsymbol{\omega}), \boldsymbol{\omega} \in \boldsymbol{\Omega} \subseteq \mathbb{C}^r\}$, the Bayesian MT-GQLRT is implemented via the following steps:

1) Obtain the optimal MT-function parameter $\boldsymbol{\omega}_{\rm opt}$ by minimizing (29) evaluated at the optimal threshold $t_{\rm opt}^{(u)}(\boldsymbol{\omega})$ (30).
2) Compute the threshold $t_{\rm opt}^{(u)}(\boldsymbol{\omega}_{\rm opt})$ using (30).
3) Apply the decision rule (20).

Similarly to implementation of the non-Bayesian MT-GQLRT, the minimization in step (1) is carried out numerically. Notice that, also here, when $\boldsymbol{\omega}$ is one-dimensional a simple line search can be implemented. Otherwise, the minimization can be performed via gradient descend or via greedy search.

[2]Notice that when $\mathbf{X}$ is a continuous random vector, this assumption satisfied almost surly [25].

## VI. Examples

In this section we illustrate the proposed MT-GQLRT and its Bayesian extension to random signal detection and deterministic signal classification, respectively. Other applications of these tests for Bayesian and non-Bayesian random signal classification are detailed in the conference papers [34], [35].

### A. Non-Bayesian MT-GQLRT: Signal detection

We consider the following signal detection problem:

$$\begin{aligned} H_0 &: \mathbf{X}_n = \mathbf{W}_n, \quad n = 1, \ldots, N, \\ H_1 &: \mathbf{X}_n = S_n \mathbf{a} + \mathbf{W}_n, \quad n = 1, \ldots, N, \end{aligned} \tag{31}$$

where $\{\mathbf{X}_n \in \mathbb{C}^p\}$, $p > 1$ is an observation process, $\{S_n \in \mathbb{C}\}$ is an i.i.d. zero-mean random signal process with *unknown* distribution, $\mathbf{a} \in \mathbb{C}^p$ is a known unit norm deterministic vector and $\{\mathbf{W}_n \in \mathbb{C}^p\}$ is an i.i.d. noise process with centered complex spherical distribution [30], i.e.,

$$\mathbf{W} \overset{d}{=} \mathbf{V}\mathbf{W} \tag{32}$$

for any unitary matrix $\mathbf{V} \in \mathbb{C}^{p \times p}$, where $\overset{d}{=}$ denotes equality in distribution. The processes $\{S_n\}$ and $\{\mathbf{W}_n\}$ are assumed to be independent. Notice that the probability distributions under each hypothesis cannot be extracted from (31) (even not up to some unknown parameters). However, as we show in the following, by specifying the MT-function in some wide class of functions partial statistical information is available through the MT-mean and the MT-covariance that are known up to some redundant constants.

In order to derive the MT-GQLRT for the considered detection problem we specify the MT-function in the set:

$$\left\{ u(\mathbf{x}) = g\left( \left\| \mathbf{P}_{\mathbf{a}}^{\perp} \mathbf{x} \right\| \right), \; g : \mathbb{R}_+ \to \mathbb{R}_+ \right\}, \tag{33}$$

where $\mathbf{P}_{\mathbf{a}}^{\perp} \triangleq \mathbf{I}_p - \mathbf{a}\mathbf{a}^H$ is the projection matrix into the null space of $\mathbf{a}$, and $\mathbf{I}_p$ is a $p \times p$ unit matrix. Assuming that condition (6) is satisfied, one can verify using (8), (10), (11), (31) and (33) that the MT-mean and the MT-covariance under the transformed probability measure $Q_k^{(u)}, k \in \{0, 1\}$ take the forms:

$$\boldsymbol{\mu}_k^{(u)} = \mathbf{0}, \qquad k = 0, 1 \tag{34}$$

and

$$\boldsymbol{\Sigma}_0^{(u)} = \boldsymbol{\Sigma}_{\mathbf{W}}^{(u)}, \qquad \boldsymbol{\Sigma}_1^{(u)} = \sigma_S^2 \mathbf{a}\mathbf{a}^H + \boldsymbol{\Sigma}_{\mathbf{W}}^{(u)}, \tag{35}$$

where $\sigma_S^2 \triangleq \mathrm{E}[|S_n|^2; P_S]$ is the signal variance, and $\boldsymbol{\Sigma}_{\mathbf{W}}^{(u)}$ is the MT-covariance of the noise. By (10), (11), (32) and (33) $\boldsymbol{\Sigma}_{\mathbf{W}}^{(u)} = r_0^{(u)} \mathbf{a}\mathbf{a}^H + r_1^{(u)} \mathbf{I}$, where $r_0^{(u)}$ and $r_1^{(u)}$ are some real constants that satisfy $r_0^{(u)}+r_1^{(u)} > 0$. The detailed algebraic manipulations showing this structure appear in [36, Sec. A]. Hence, by substituting (34) and (35) into (18) the resulting test statistic after subtraction of the observation-independent constant $c_1^{(u)} \triangleq -\log\left(1+\sigma_S^2/(r_0^{(u)}+r_1^{(u)})\right)$ followed by normalization by the positive observation-independent constant $c_2^{(u)} \triangleq \sigma_S^2/((r_0^{(u)}+r_1^{(u)})(r_0^{(u)}+r_1^{(u)}+\sigma_S^2))$ is given by:

$$T_u' \triangleq \frac{T_u - c_1^{(u)}}{c_2^{(u)}} = \mathbf{a}^H \hat{\mathbf{C}}^{(u)} \mathbf{a} = \sum_{n=1}^{N} \hat{\varphi}_u(\mathbf{X}_n) |\mathbf{a}^H \mathbf{X}_n|^2, \tag{36}$$

where $\hat{\mathbf{C}}^{(u)} \triangleq \hat{\mathbf{\Sigma}}^{(u)} + \hat{\boldsymbol{\mu}}^{(u)} \hat{\boldsymbol{\mu}}^{(u)H}$. Notice that when the vector $\mathbf{a}$ represents a steering vector of a sensor array [37], the test statistic in (37) is a measure transformed version of Bartlett's beamformer [37]. Finally, by (20) the MT-GQLRT is given by

$$T_u' \underset{H_0}{\overset{H_1}{\gtrless}} t', \tag{37}$$

where $t' \triangleq (t - c_1^{(u)})/c_2^{(u)}$.

Under the considered settings, it can be shown that the conditions stated in Theorem 1 are satisfied. Since the noise vector $\mathbf{W}$ has a centered spherical distribution, then it must obey the following stochastic representation [30]:

$$\mathbf{W} \overset{d}{=} \sigma_{\mathbf{W}} Y \mathbf{U}, \tag{38}$$

where $\sigma_{\mathbf{W}} \in \mathbb{R}_{++}$ is a scale parameter, $Y \in \mathbb{R}_{+}$ is a real non-negative random variable, called modular variate, and $\mathbf{U} \in \mathbb{C}^p$ is a random vector, that is statistically independent of $Y$, with uniform distribution on the unit complex $p$-sphere. Therefore, by (33)-(35) and (38) the resulting asymptotic power (23) at a given asymptotic size $\alpha_u = \alpha$ takes the form:

$$\beta_u^{(\alpha)} = Q\left(\frac{\sqrt{G_1} Q^{-1}(\alpha) - \sqrt{N}\sigma_S^2}{\sqrt{G_2}}\right), \tag{39}$$

where $G_1 \triangleq \mathrm{E}[g^2(\tilde{Y}\sqrt{1-B})(|S|^2 - \sigma_S^2 + \tilde{Y}^2 B - h)^2; P_{S,Y,B}]$, $G_2 \triangleq \mathrm{E}[g^2(\tilde{Y}\sqrt{1-B})(\tilde{Y}^2 B - h)^2; P_{Y,B}]$, $h \triangleq \mathrm{E}[g(\tilde{Y}\sqrt{1-B})\tilde{Y}^2 B; P_{Y,B}]$, $\tilde{Y} \triangleq \sigma_{\mathbf{W}} Y$, and $B \sim \mathrm{Beta}(1, p-1)$. Furthermore, its empirical estimate is given by:

$$\hat{\beta}_u^{(\alpha)} = Q\left(\frac{\tilde{\eta}_0^{(u)} - \tilde{\eta}_1^{(u)} + \sqrt{\tilde{\lambda}_0^{(u)}} Q^{-1}(\alpha)}{\sqrt{\tilde{\lambda}_1^{(u)}}}\right), \tag{40}$$

where

$$\tilde{\eta}_k^{(u)} \triangleq \frac{\hat{\eta}_k^{(u)} - c_1^{(u)}}{c_2^{(u)}} = \sum_{n=1}^{N_k} \hat{\varphi}_u(\mathbf{X}_n^{(k)}) \left|\mathbf{a}^H \mathbf{X}_n^{(k)}\right|^2,$$

$$\tilde{\lambda}_k^{(u)} \triangleq \frac{\hat{\lambda}_k^{(u)}}{(c_2^{(u)})^2} = \frac{N_k}{N} \sum_{n=1}^{N_k} \hat{\varphi}_u^2(\mathbf{X}_n^{(k)}) \left(|\mathbf{a}^H \mathbf{X}_n^{(k)}|^2 - \tilde{\eta}_k^{(u)}\right)^2,$$

and $c_1^{(u)}$ and $c_2^{(u)}$ are defined above (36). We note that by (24) the threshold of the decision rule (37) that corresponds to a constant asymptotic test-size $\alpha$ is given by:

$$t' = \tilde{\eta}_0^{(u)} + \sqrt{\tilde{\lambda}_0^{(u)}} Q^{-1}(\alpha). \tag{41}$$

In order to mitigate the effect of outliers and involve higher-order statistical moments, we specify the MT-function in a subset of (33) that is comprised of zero-centred Gaussian functions parametrized by a width parameter $\omega$, i.e.,

$$u_G(\mathbf{x}; \omega) = \exp\left(-\left\|\mathbf{P}_{\mathbf{a}}^{\perp}\mathbf{x}\right\|^2 / \omega^2\right), \quad \omega \in \mathbb{R}_{++}. \tag{42}$$

Notice that the Gaussian MT-function (42) does not shrink outliers in the direction of the vector $\mathbf{a}$ and does not satisfy the B-robustness condition (15) when the observations are proportional to $\mathbf{a}$. However, this shrinkage does occur over a sufficiently large subset of $\mathbb{C}^p$, guaranteeing robustness of the empirical MT-mean and MT-covariance with high probability. To see this, define the set $\mathcal{B}_\epsilon \triangleq \{\mathbf{y} \in \mathbb{C}^p : \left|\mathbf{a}^H\mathbf{y}\right|^2 / \|\mathbf{y}\|^2 \leq 1 - \epsilon\}$, where $\epsilon > 0$ is some fixed small positive constant. Clearly, $u_G(\mathbf{y}; \omega) \leq \exp\left(-\frac{\epsilon\|\mathbf{y}\|^2}{\omega^2}\right)$ for any $\mathbf{y} \in \mathcal{B}_\epsilon$ and for any fixed $\omega$. Therefore, since $\exp\left(-\epsilon\|\mathbf{y}\|^2/\omega^2\right)$ and $\|\mathbf{y}\|^2 \exp\left(-\epsilon\|\mathbf{y}\|^2/\omega^2\right)$ are bounded over $\mathbb{C}^p$, the MT-function (42) must satisfy condition (15) over $\mathcal{B}_\epsilon$. Finally, since $P(\mathcal{B}_\epsilon) \approx 1$ for sufficiently small $\epsilon$ we conclude that the empirical MT-mean and MT-covariance, comprising (36), are robust to outliers with high probability. Moreover, similarly to Proposition 4 in [17], it can be shown that for any fixed width parameter $\omega$, the influence functions of the empirical MT-mean and MT-covariance, comprising (36), approach to zero over the set $\mathcal{B}_\epsilon$ as the outlier norm approaches to infinity. Thus, we conclude that the MT-function (42) also results in rejection of large norm outliers with high probability. Furthermore, notice that the Gaussian MT-function (42) is parameterized by only one scalar parameter $\omega$. This leads to a simple line search based optimization of the empirical asymptotic power (40) w.r.t. $\omega$.

In the following simulation examples we evaluate the detection performance of the MT-GQLRT as compared to the omniscient LRT, the standard GQLRT (3), a robust GQLRT extension, a density-estimator plug-in detector, the NSDD-GLRT [38], and a support vector machine (SVM) [39].

**Robust GQLRT extension:** Under the considered detection problem (31) one can verify using (3) that the test-statistic of the GQLRT reduces to $T_{\text{GQLRT}} = \mathbf{a}^H\hat{\mathbf{C}}\mathbf{a}$, where $\hat{\mathbf{C}} \triangleq \sum_{n=1}^{N} \mathbf{X}_n\mathbf{X}_n^H$ is the non-robust sample correlation matrix. Hence, a robust extension of the GQLRT can be obtained by applying GQLRT after passing the data through a zero-memory non-linear (ZMNL) function that suppresses outliers by clipping the amplitude of the observations. This GQLRT extension is called here ZMNL-GQLRT. We use the same ZMNL preprocessing approach that has been applied in [40].

**Density-estimator plug-in detector:** When training sequences $\mathbf{X}_n^{(k)}$, $n = 1, \ldots, N_k$, $k = 0, 1$ from $P_0$ and $P_1$ are available to estimate the densities $f_1$ and $f_0$, a density-plug-in approach is natural. One can approximate the LRT by estimating the probability density functions under each hypothesis, and perform the test $\sum_{n=1}^{N} \log(\hat{f}_1(\mathbf{X}_n)/\hat{f}_0(\mathbf{X}_n)) \gtrless_{H_0}^{H_1} t$, where $\hat{f}_1(\cdot)$ and $\hat{f}_0(\cdot)$ are estimates of $f_1(\cdot)$ and $f_0(\cdot)$. Here, we consider parametric density estimation based on the generalized Gaussian distribution (GGD). More specifically, we assume that under each hypothesis the observations obey a GGD [30] with zero location parameter and a structured scatter matrix that is proportional to the covariance under (31). The shape parameter of the GGD distribution and the parameters of the scatter matrix under each hypothesis were estimated via straightforward iterative maximum-likelihood (ML) estimator. Exact implementation details of the ML-estimator appear in [36, Sec. B]. The detector based on this approach is called here GGD-QLRT. Note that the GGD-QLRT belongs to the class of non-Gaussian quasi likelihood ratio tests that replace the true likelihoods with hypothesized parametric likelihoods.

**NSDD-GLRT [38]:** The NSDD-GLRT is a robust generalized likelihood ratio test (GLRT) detector, which assumes that the signal samples in (31) are deterministic unknown and that the noise samples are zero-mean normally distributed with unknown variances.

**SVM based detector:** In this example, the separation between the hypotheses is non-linear. Therefore, a kernel SVM [39] was implemented that applies SVM to high-dimensional non-linear transformations of the observation vectors, that map them into some reproducing kernel Hilbert spaces [41]. Exact implementation details of the SVM based detector appear in [36, Sec. C].

In all simulation examples, the signal $S_n$ in (31) is considered to be a BPSK signal with power $\sigma_S^2$. The vector $\mathbf{a} \triangleq \frac{1}{\sqrt{p}}[1, e^{-i\pi \sin(\vartheta)}, \ldots, e^{-i\pi(p-1)\sin(\vartheta)}]^T$ represents a steering vector of $p = 8$ elements uniform linear array with half wavelength spacing corresponding to a far-field narrow band signal with azimuthal angle of arrival (AOA) $\vartheta = \pi/3$ [Rad]. We considered two types of noise distributions with zero location parameter and isotropic dispersion $\sigma_{\mathbf{W}}^2 \mathbf{I}_p$: 1) Gaussian and 2) $\epsilon$-contaminated Gaussian noise [30] under which $\mathbf{W} \stackrel{d}{=} \sigma_{\mathbf{W}}^2 A \mathbf{Z}$, where $A$ is a binary random variable satisfying $A = 1$ w.p. $1 - \epsilon$ and $A = \delta$ w.p. $\epsilon$, and $\mathbf{Z} \sim \mathcal{CN}(\mathbf{0}, \mathbf{I}_p)$. The parameters $\epsilon$ and $\delta$ that control the heaviness of the noise tails were set to 0.25 and 10, respectively. Notice that in the context of the stochastic representation (38) $Y = A\|\mathbf{Z}\|$ and $\mathbf{U} = \mathbf{Z}/\|\mathbf{Z}\|$.

For each noise type we performed two simulations. In the first one, we compared the asymptotic power (39) to its empirical estimate (40) as a function of $\omega$ for a fixed asymptotic test size $\alpha = 10^{-3}$ and sample size $N = 300$. The empirical asymptotic power (40) was obtained using two i.i.d. training sequences from $P_0$ and $P_1$ containing $N_0 = N_1 = 3 \times 10^4$ samples. The signal-to-noise-ratio (SNR), defined here as SNR $\triangleq 10\log_{10}\sigma_S^2/\sigma_W^2$ was set to $-5$ [dB]. Observing Figs. 1(a) and 2(a), one sees that due to the consistency of (40) the compared quantities are very close. This illustrates the reliability of the empirical asymptotic power for optimal choice of the MT-function parameter, as discussed in subsection IV-C.

In the second simulation, we compared the empirical power of the proposed test to the empirical powers obtained by the other compared tests versus SNR, samples size $N$, and test size (ROC curve). For each type of comparison, we also report the optimal asymptotic power of the MT-GQLRT that is obtained by maximizing (39) w.r.t. the width parameter $\omega \in \Omega \triangleq [1, 100]$ of the Gaussian MT-function (42).

The MT-GQLRT was implemented in two manners:

1) **Optimal implementation requiring training sequences:** Here, the proposed test (37) was implemented in accordance to the steps detailed in Subsection IV-D, that involve two training sequences. More specifically, the empirical asymptotic power (40) was computed using two training sequences of size $N_0 = N_1 = 3 \times 10^4$. The optimal Gaussian MT-function parameter $\omega_{\rm opt}$ was obtained by minimizing (40) over $K_\Omega = 100$ equally spaced grid points of the interval $\Omega$ defined above. The threshold was determined directly from (41). This optimal implementation will be called "MT-GQLRT$_{\rm opt}$".

2) **Suboptimal implementation not requiring training sequences:** Here, implementation steps (2) and (3) detailed in Subsection IV-D are modified. More specifically, suboptimal selection of the width parameter $\omega$ is carried out via data-driven procedure that exploits only the test sequence itself. This procedure is described in Appendix F. Furthermore, the threshold is determined via Monte-Carlo simulations as detailed below (and not via (41)). This suboptimal implementation will be called "MT-GQLRT$_{\rm sub}$".

In the GGD-QLRT and the SVM we used the same training sequences of size $N_0 = N_1 = 3 \times 10^4$ that were used by the MT-GQLRT$_{\rm opt}$. For all compared tests, except the MT-GQLRT$_{\rm opt}$, Monte-Carlo simulations were performed in-order to determine the threshold value corresponding to a fixed test-size $\alpha$, by estimating the $1-\alpha$ percentile of the test-statistic under the null hypothesis. The Monte-Carlo simulations were carried out using $M = 10^5$ i.i.d. training sequences of size $N$ from $P_0$. The empirical power curves were obtained using $10^5$ Monte-Carlo simulations. The detection performance versus SNR, sample size and the test size are depicted in Figs. 1(b)–1(d) for the Gaussian noise and in Figs. 2(b)–2(d) for the non-Gaussian noise. The power versus SNR was evaluated for a fixed test size equal to $10^{-3}$ and $N = 300$ i.i.d. observations. The power versus sample size was evaluated for a fixed test size equal to $10^{-3}$, and SNR $= -10$ [dB] for the Gaussian noise, and SNR $= -9$ [dB] for the non-Gaussian noise. The power versus test size (ROC curve) was evaluated for $N = 300$ i.i.d. observations and SNR $= -10$ [dB]. Observing Figs. 1(b)–1(d), one can notice that the MT-GQLRT$_{\rm opt}$, GQLRT and GGD-QLRT attain similar performance. The MT-GQLRT$_{\text{sub}}$ performs similarly to the NSDD-GLRT. The agreement between the MT-GQLRT$_{\text{opt}}$ and the standard GQLRT is an outcome of the fact that the MT-GQLRT$_{\text{opt}}$ approaches the GQLRT as the width parameter of the Gaussian MT-function (42) approaches infinity. Observing Figs. 2(b)−2(d), one sees that for the non-Gaussian noise, the MT-GQLRT$_{\text{opt}}$ outperforms the non-robust GQLRT and all other robust alternatives. It also attains detection performance that is significantly closer to that of the LRT that, unlike the MT-GQLRT$_{\text{opt}}$, requires complete knowledge of the likelihood function under each hypothesis. Furthermore, one sees that although the MT-GQLRT$_{\text{sub}}$, which does not involve training sequences for selection of $\omega$, is inferior as compared to the MT-GQLRT$_{\text{opt}}$, it outperforms all other robust GQLRT alternatives.

A general asymptotic computational load (ACL) analysis (for the considered detection problem) is reported in Table I. Notice that both MT-GQLRT$_{\text{opt}}$ and MT-GQLRT$_{\text{sub}}$ have the same ACL for detection as the standard GQLRT. Also note that the ACL of the MT-GQLRT$_{\text{opt}}$ due to parameter tuning (optimization of the width parameter of the Gaussian MT-function, which is performed via simple line search) is linear in the sample size, dimension and number of grid points taken over $\Omega$. Furthermore, the ACL of the MT-GQLRT$_{\text{sub}}$ due to parameter tuning is linear in sample size and dimension. Although the ACLs of the MT-GQLRT$_{\text{opt}}$ due to parameter tuning and detection are similar to those of the GGD-QLRT, the MT-GQLRT$_{\text{opt}}$ outperforms the GGD-QLRT as discussed above.

Finally, additional analysis of the compared detectors for small sample size is provided in [36, Sec. F]. Furthermore, a modified scale-invariant version of the proposed test (37) is presented in [36, Sec. G].

### *B. Bayesian MT-GQLRT: Signal classification*

We consider the following Bayesian signal classification problem:

$$\begin{aligned} H_0 &: \mathbf{X}_n = \mathbf{a}_0 + \mathbf{W}_n, \;\; n = 1, \ldots, N, \\ H_1 &: \mathbf{X}_n = \mathbf{a}_1 + \mathbf{W}_n, \;\; n = 1, \ldots, N, \end{aligned} \tag{43}$$

with known a-priori probabilities $\pi_0$ and $\pi_1$. Here, $\{\mathbf{X}_n \in \mathbb{C}^p\}$, $p > 2$ is an observation process, and $\mathbf{a}_0, \mathbf{a}_1 \in \mathbb{C}^p$ are known deterministic vector signals. Similarly to the detection problem in the previous subsection, we assume that $\{\mathbf{W}_n \in \mathbb{C}^p\}$ is a spherically contoured i.i.d. noise process that obeys the stochastic representation (38). Generally, this is a location parameter classification problem [42] when multiple instances from each class are available [42], [43].

[3]Since in this example the MT-function width parameter $\omega$ is tuned through a simple line search, then by (40) and (41) the threshold can be computed directly from the quantities in the tuning process.

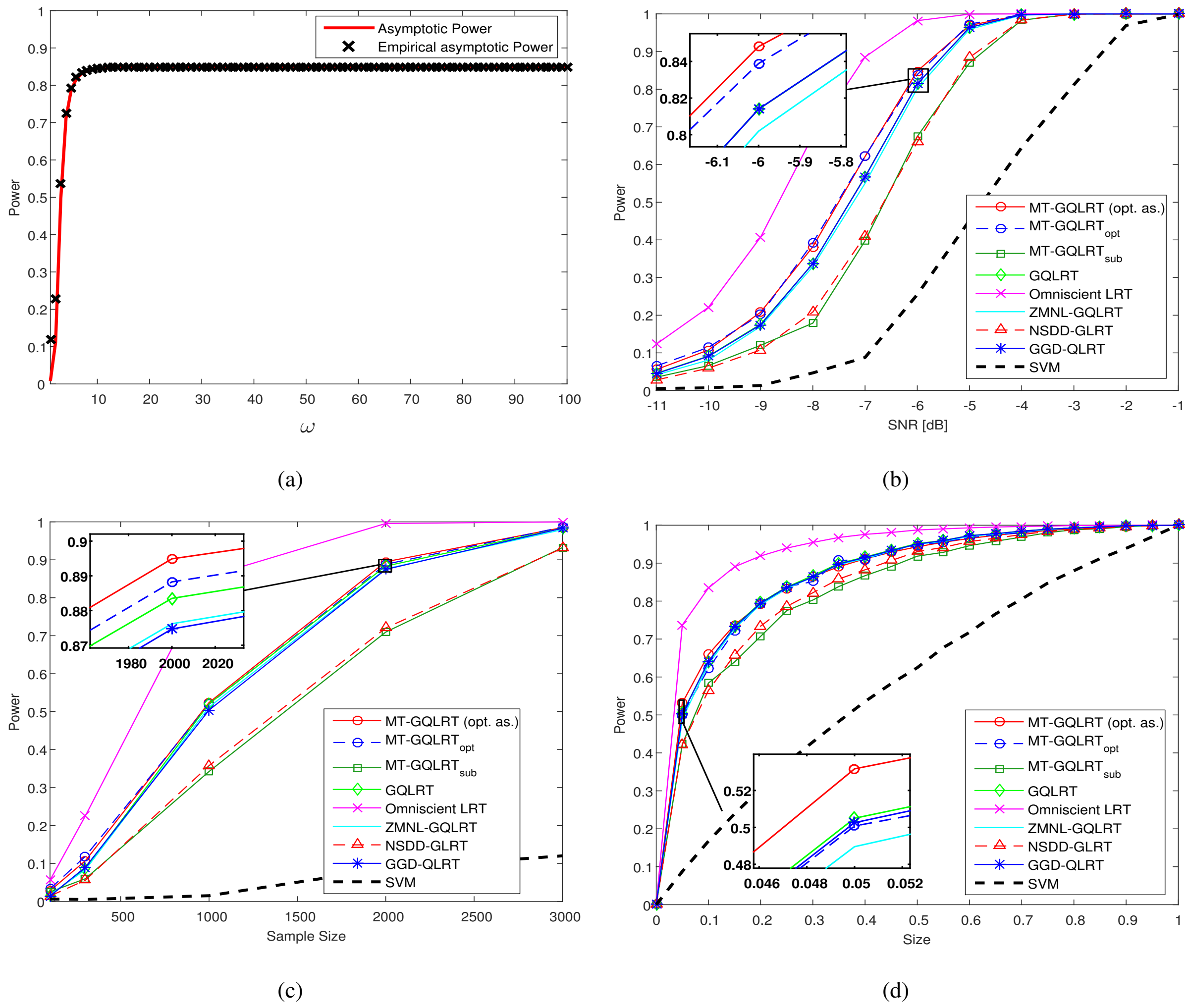

Fig. 1. **Signal detection in Gaussian noise:** (a) Asymptotic power predicted by theory (39) and its empirical estimate (40) versus the width parameter $\omega$ of the Gaussian MT-function (42). (b) + (c) + (d) Optimal asymptotic power of the MT-GQLRT, and the empirical powers of the MT-GQLRT$_{\text{opt}}$ and the MT-GQLRT$_{\text{sub}}$ as a function of (b) SNR, (c) sample size and (d) test size as compared to the empirical powers of the GQLRT, ZMNL-GQLRT, NSDD-GLRT, GGD-QLRT, SVM and the omniscient LRT.

In order to derive the MT-GQLRT for the Bayesian decision problem (43) we specify the MT-function in the set:

$$\left\{u\left(\mathbf{x}\right)=g\left(\left\|\mathbf{P}_{\mathbf{A}}^{\perp}\mathbf{x}\right\|\right),\ g:\mathbb{R}_{+}\rightarrow\mathbb{R}_{+}\right\}, \tag{44}$$

where $\mathbf{A}\triangleq[\mathbf{a}_0,\mathbf{a}_1]$ and $\mathbf{P}_{\mathbf{A}}^{\perp}$ is the projection matrix into the null space of $\mathbf{A}$. Similarly to the signal detection problem in the previous subsection, assuming that condition (6) is satisfied under the conditional probability measure $P$, one can verify using (8), (10), (11), (43) and (44) that the conditional MT-mean

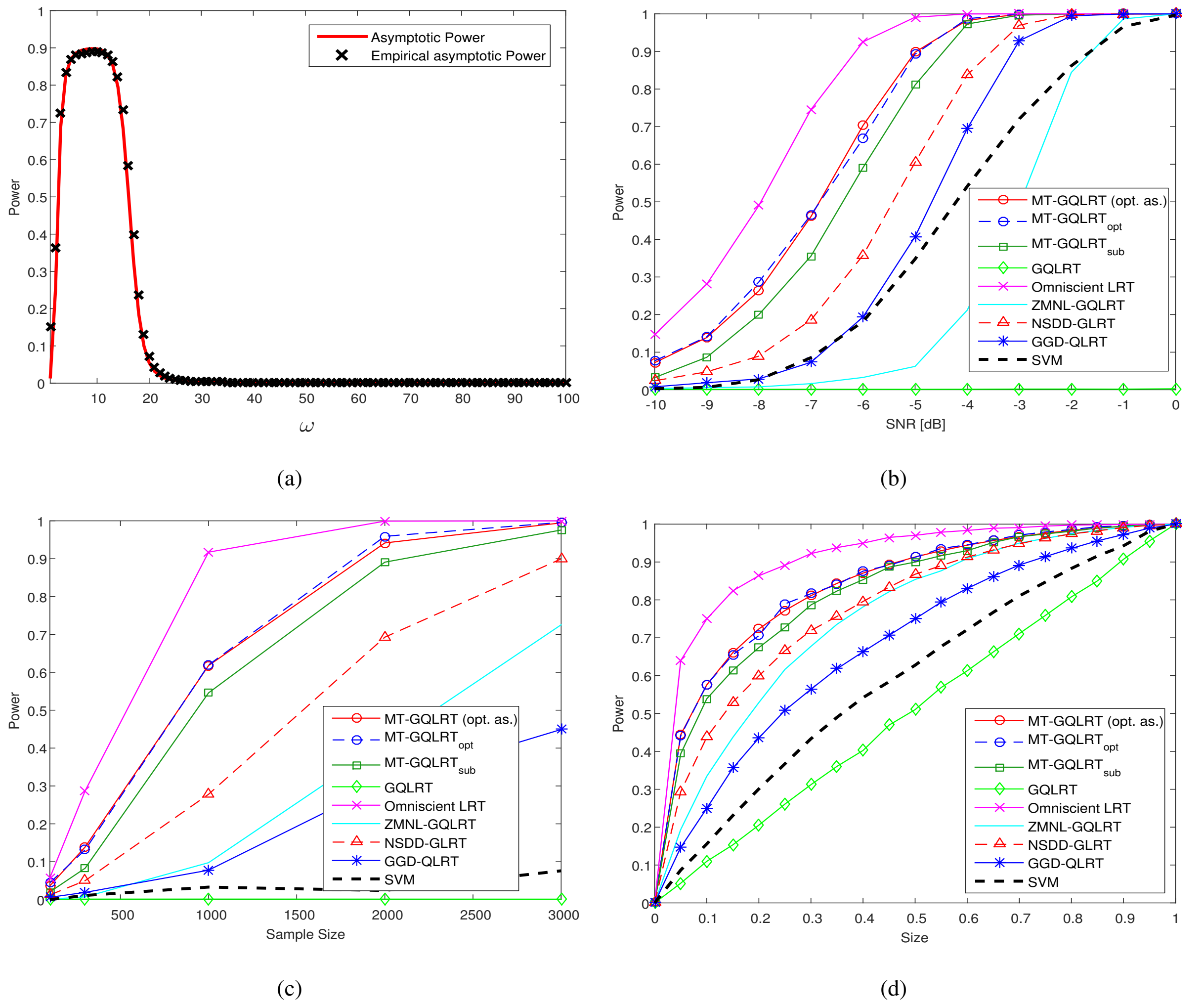

Fig. 2. **Signal detection in non-Gaussian noise:** (a) Asymptotic power predicted by theory (39) and its empirical estimate (40) versus the width parameter $\omega$ of the Gaussian MT-function (42). (b) + (c) + (d) Optimal asymptotic power of the MT-GQLRT, and the empirical powers of the MT-GQLRT$_{\text{opt}}$ and the MT-GQLRT$_{\text{sub}}$ as a function of (b) SNR, (c) sample size and (d) size as compared to the empirical powers of the GQLRT, ZMNL-GQLRT, NSDD-GLRT, GGD-QLRT, SVM and the omniscient LRT.

and MT-covariance satisfy the following properties:

$$\boldsymbol{\mu}_k^{(u)} = \mathbf{a}_k \qquad k = 0, 1 \tag{45}$$

and

$$\boldsymbol{\Sigma}_k^{(u)} = r_0^{(u)} \mathbf{P}_{\mathbf{A}} + r_1^{(u)} \mathbf{I} \qquad k = 0, 1 \tag{46}$$

where $\mathbf{P}_{\mathbf{A}}$ is the projection matrix onto the range space of $\mathbf{A}$, and $r_0^{(u)}$ and $r_1^{(u)}$ are some constants that satisfy $r_0^{(u)} + r_1^{(u)} > 0$. Hence, by substituting (45) and (46) into (18) the resulting test statistic

TABLE I

*Signal detection:* ASYMPTOTIC COMPUTATIONAL COMPLEXITY (FLOPS). NOTATION: $p$ IS THE DIMENSION OF THE OBSERVATION VECTORS. $N$ DENOTES THE SAMPLE SIZE. $N_0$ AND $N_1$ ARE THE SIZES OF THE TRAINING SEQUENCES FROM $H_0$ AND $H_1$, RESPECTIVELY. $K_\Omega$ DENOTES THE NUMBER OF GRID POINTS OF THE $\Omega$-AXIS (THE WIDTH PARAMETER SPACE OF THE GAUSSIAN MT-FUNCTION (42)). $I$ DENOTES NUMBER OF ITERATIONS. $M$ DENOTES THE NUMBER OF TRIALS FOR ESTIMATING THE $1-\alpha$ PERCENTILE OF THE TEST STATISTIC.

| Method | Parameter tuning | Threshold calculation | Detection |
|---|---|---|---|
| MT-GQLRT$_{\text{opt}}$ | $O((N_0+N_1)pK_\Omega)$ | $O(1)$[3] | $O(Np)$ |
| MT-GQLRT$_{\text{sub}}$ | $O(Np)$ | $O(MNp)$ | $O(Np)$ |
| GQLRT | – | $O(MNp)$ | $O(Np)$ |
| ZMNL-GQLRT | – | $O(MNp)$ | $O(Np)$ |
| NSDD-GQLRT | – | $O(MNp)$ | $O(Np)$ |
| GGD-QLRT | $O((N_0+N_1)pI)$ | $O(MNp)$ | $O(Np)$ |
| SVM | $O((N_0+N_1)pI)$ | $O(MNp)$ | $O(Np)$ |

after subtraction of the observation-independent constant $c_1^{(u)} \triangleq r_1^{(u)}(\|\mathbf{a}_0\|^2 - \|\mathbf{a}_1\|^2)/(r_1^{(u)}(r_0^{(u)} + r_1^{(u)}))$ followed by normalization by the positive observation-independent factor $c_2^{(u)} \triangleq 2/(r_0(\omega) + r_1(\omega))$ is given by:

$$T'_u \triangleq \frac{T_u - c_1^{(u)}}{c_2^{(u)}} = \text{Re}\left\{(\mathbf{a}_1 - \mathbf{a}_0)^H \hat{\boldsymbol{\mu}}^{(u)}\right\} \underset{H_0}{\overset{H_1}{\gtrless}} t', \tag{47}$$

where $t' \triangleq (t - c_1^{(u)})/c_2^{(u)}$.

We choose the loss coefficients $L_{10} = L_{01} = 1$, under which the asymptotic Bayes risk (28) reduces to the probability of error [1]. In this case, using (28), (38), (45) and (46), it can be shown that the asymptotic minimum probability of error w.r.t. the threshold parameter takes the form:

$$P_e^{(u)} = \sum_{k=0}^{1} \pi_k Q\left(G_u + (-1)^k \frac{1}{2G_u} \log \frac{\pi_0}{\pi_1}\right), \tag{48}$$

where $G_u \triangleq \frac{\sqrt{N}\|\mathbf{a}_1 - \mathbf{a}_0\| \text{E}[g(\tilde{Y}\sqrt{1-C}); P_{Y,C}]}{\sqrt{2\text{E}[\tilde{Y}^2 C g^2(\tilde{Y}\sqrt{1-C}); P_{Y,C}]}}$, $\tilde{Y} \triangleq \sigma_{\mathbf{W}} Y$, $C \triangleq 2/(2 + (p-2)Z)$, $Z \sim F(2p-4, 4)$, and $F(k,l)$ denotes an F-distribution with $k$ and $l$ degrees of freedom. Moreover, by (29) and (30) the empirical estimate of (48) is given by:

$$\hat{P}_e^{(u)} = \sum_{k=0}^{1} \pi_k Q\left(\frac{\tilde{t}_{\text{opt}}^{(u)} - \tilde{\eta}_k^{(u)}}{\sqrt{\tilde{\lambda}_k^{(u)}}}\right), \tag{49}$$

where

$$\tilde{\eta}_k^{(u)} \triangleq \frac{\hat{\eta}_k^{(u)} - c_1^{(u)}}{c_2^{(u)}} = \sum_{n=1}^{N} \hat{\varphi}_u(\mathbf{X}_n^{(k)}) \, \mathrm{Re}\{(\mathbf{a}_1 - \mathbf{a}_0)^H \mathbf{X}_n^{(k)}\},$$

$$\tilde{\lambda}_k^{(u)} \triangleq \frac{\hat{\lambda}_k^{(u)}}{(c_2^{(u)})^2} = \frac{N_k}{N} \sum_{n=1}^{N_k} \hat{\varphi}_u^2(\mathbf{X}_n^{(k)}) \left(\mathrm{Re}\{(\mathbf{a}_1 - \mathbf{a}_0)^H \mathbf{X}_n^{(k)}\} - \tilde{\eta}_k^{(u)}\right)^2,$$

$k = 0, 1$, and the optimal threshold $\tilde{t}_{\text{opt}}^{(u)}$ is obtained from (30) by replacing $\hat{\eta}_k^{(u)}$ and $\hat{\lambda}_k^{(u)}$ with $\tilde{\eta}_k^{(u)}$ and $\tilde{\lambda}_k^{(u)}$. As discussed in Subsection V-B, the empirical error probability (49) will be used for optimal choice of the MT-function parameters.

Similarly to the detection problem in the previous subsection, in order to mitigate the effect of outliers and involve higher-order statistical moments, we specify the MT-function in a subset of (44) that is comprised of zero-centred Gaussian functions parametrized by a width parameter $\omega$, i.e.,

$$u_G(\mathbf{x}; \omega) = \exp\left(-\left\|\mathbf{P}_{\mathbf{A}}^{\perp}\mathbf{x}\right\|^2 / \omega^2\right), \quad \omega \in \mathbb{R}_{++}. \tag{50}$$

Similarly to the signal detection example, it can be shown that the resulting empirical MT-mean that comprise the test-statistic is B-robust and rejects large norm outliers with high probability.

In the following simulation examples we compare the classification performance of the MT-GQLRT (47) to the Bayesian versions of the omniscient LRT, the standard GQLRT, other robust GQLRT extensions, a density-estimator plug-in classifier and SVM.

**Robust GQLRT extensions:** Under the classification problem (43) one can verify that the test-statistic of the GQLRT reduces to $T_{\text{GQLRT}} = \mathrm{Re}\{(\mathbf{a}_1 - \mathbf{a}_0)^H \hat{\boldsymbol{\mu}}\}$, where $\hat{\boldsymbol{\mu}}$ is the standard SMV. Hence, other robust alternatives to the GQLRT can be obtained by replacing the non-robust SMV with robust location estimates, namely, the median estimator, and Tukey's bi-square M-estimator [29]. The robust GQLRT extension that uses the median estimator is called here Median-GQLRT. Tukey's bi-square M-estimator involves a tunning parameter $c$ that controls the shrinkage level of outliers. Here, this tuning parameter was determined in two different manners: a) Training-sequences-free approach: Here, the tuning parameter was set to guarantee an asymptotic relative efficiency of 95% of the location estimate, relative to the Cramér-Rao lower bound [1] under nominal Gaussian distribution. b) Training-sequences-based approach: This approach assumes that training sequences from $P_0$ and $P_1$ are available. Similarly to the MT-GQLRT, the optimal tuning parameter is the one that minimizes an empirical estimate of the corresponding asymptotic probability of error. These two selection procedures of the tuning parameter result in two GQLRT extensions that are called here Tukey-GQLRT$_{\text{sub}}$ and Tukey-GQLRT$_{\text{opt}}$, respectively. The exact implementation details appear in [36, Sec. D].

**Density-estimator plug-in classifier:** Similarly to the signal detection example in the previous subsection, given training sequences $\mathbf{X}_n^{(k)}$, $n = 1, \ldots, N_k$, $k = 0, 1$ from $P_0$ and $P_1$, one can approximate the LRT by estimating the conditional probability density functions under each hypothesis. Following (43), we consider a parametric set of distributions with *known* location parameters, i.e., $\mathbf{a}_0$ under $H_0$ and $\mathbf{a}_1$ under $H_1$, and spherical scatter matrices, with unknown scale parameter. Similarly to the signal detection example, we chose the elliptical family of GGDs [30]. The parameters of this distribution (shape and scale) were estimated using a straightforward iterative ML estimator. Exact implementation details of the ML estimator appear in [36, Sec. E]. The classifier based on this approach is called here GGD-QLRT.

**SVM based classifier:** Similarly to the signal detection example, an SVM classifier was trained using two training sequences (one from each hypothesis). Since in this example the separation between the hypotheses is linear, a linear SVM was implemented. We applied the same SVM based decision rule as in Eq. (S-8) in [36]. The difference, is that the threshold was selected (via Monte-Carlo simulations) by minimizing the empirical probability of error.

In all examples, the vectors $\mathbf{a}_0$ and $\mathbf{a}_1$ were set to $\mathbf{a}_k \triangleq s_k[1, e^{-i\pi/p}, \ldots, e^{-i\pi(p-1)/p}]^T$, $k = 0, 1$, where $s_0 = 5$, $s_1 = 5.25$ and $p = 10$. The a-priori probabilities were set to $\pi_0 = 0.6$ and $\pi_1 = 0.4$. We considered two types of noise distributions with zero location parameter and isotropic dispersion $\sigma_{\mathbf{W}}^2 \mathbf{I}_p$: 1) Gaussian and 2) $t$-distributed noise [30] with $\lambda = 0.2$ degrees of freedom.

Similarly to the signal detection example, for each noise type we performed two simulations. In the first simulation example, we compared the asymptotic probability of error (48) to its empirical estimate (49) as a function of $\omega$ for sample size of $N = 300$. The empirical asymptotic probability of error (49) was obtained using two i.i.d. training sequences from $P_0$ and $P_1$ containing $N_0 = N_1 = 3 \times 10^4$ samples. The SNR, defined in this example as $\text{SNR} \triangleq 10 \log_{10} (\|\mathbf{a}_0 - \mathbf{a}_1\|)^2 / \sigma_{\mathbf{W}}^2$, was set to $-18$ [dB]. Observing Figs. 3(a) and 4(a), one sees that due to the consistency of (49) the compared quantities are very close. This illustrates the reliability of the empirical asymptotic Bayes risk for optimal choice of the MT-function parameter, as discussed in subsection V-B.

In the second simulation, we compared the empirical probability of error of the proposed test to the empirical probability of errors obtained by the other compared tests versus SNR and samples size $N$. For each type of comparison, we also report the optimal asymptotic probability of error of the MT-GQLRT that is obtained by minimizing (48) w.r.t. the width parameter $\omega \in \Omega \triangleq [1, 100]$ of the Gaussian MT-function (50).

Similarly to previous application example, the Bayesian MT-GQLRT was implemented in two different manners:

1) **Optimal implementation requiring training sequences:** Here, the proposed test (47) was implemented in accordance to the steps detailed in Subsection V-C that involve two training sequences. The empirical asymptotic probability of error (49) was computed using two training sequences of size $N_0 = N_1 = 3 \times 10^4$. The optimal Gaussian MT-function parameter $\omega_{\text{opt}}$ was obtained by minimizing (49) over $K_\Omega = 100$ grid points of the interval $\Omega$ defined above. The threshold value was determined as described below (49) and it was evaluated at $\omega_{\text{opt}}$. This optimal implementation is called here "MT-GQLRT$_{\text{opt}}$".

2) **Suboptimal implementation not requiring training sequences:** Here, similarly to the signal detection example, the width parameter of the Gaussian MT-function (50) was selected via data-driven procedure that involves only the test sequence itself. This procedure is described in Appendix G. The threshold parameter is obtained via Monte-Carlo simulations as discussed below. This suboptimal implementation is called "MT-GQLRT$_{\text{sub}}$".

The Tukey-GQLRT$_{\text{opt}}$, GGD-QLRT and SVM were implemented using the same training sequences of size $N_0 = N_1 = 3\times 10^4$ used by the MT-GQLRT$_{\text{opt}}$. For all compared tests, except the MT-GQLRT$_{\text{opt}}$ and Tukey-GQLRT$_{\text{opt}}$, Monte-Carlo simulations were performed in-order to determine the threshold value that minimizes the probability of error. These Monte-Carlo simulations were carried out using $M = 10^5$ i.i.d. training sequences of size $N$ from each hypothesis. Similarly to the MT-GQLRT$_{\text{opt}}$, the threshold for the Tukey-GQLRT$_{\text{opt}}$ was determined by minimizing the empirical asymptotic probability of error w.r.t. the threshold as described in [36, Sec. D]. The empirical probability of error curves were obtained using $10^5$ Monte-Carlo simulations. The SNR and sample size are used to index the classification performances as depicted in Figs. 3(b) and 3(c) for the Gaussian noise and in Figs. 4(b) and 4(c) for the $t$-distributed noise. The probability of error versus SNR was evaluated for $N = 300$ i.i.d. observations, and the probability of error versus sample size was evaluated for $\text{SNR} = -22$ [dB]. Observing Figs. 3(b) and 3(c), one can notice that, as expected, when the noise is Gaussian, the MT-GQLRT$_{\text{opt}}$ achieves the LRT performance and outperforms the Median-GQLRT, the Tukey-GQLRT$_{\text{sub}}$ and the SVM. Observing Figs. 4(b) and 4(c), one sees that for the $t$-distributed noise, the MT-GQLRT$_{\text{opt}}$ outperforms the non-robust Bayesian GQLRT and the other compared methods and attains classification performance that are much closer to those obtained by the Bayesian LRT that, unlike the MT-GQLRT$_{\text{opt}}$, requires complete knowledge of the conditional likelihood function under each hypothesis. The MT-GQLRT$_{\text{sub}}$ (which do not involve training sequences for selection of $\omega$) outperforms any detector that do not use training sequences. Notice that, although the Tukey-GQLRT$_{\text{opt}}$ performs similarly to the MT-GQLRT$_{\text{opt}}$, it has a significantly higher computational complexity as described in Table II below. Furthermore, unlike the MT-GQLRT$_{\text{opt}}$, it involves an iterative process for estimating the location vector parameter.

A general asymptotic computational load (ACL) analysis (for the considered classification problem) is reported in Table II. Similarly to the signal detection problem, notice that the MT-GQLRT$_{\text{opt}}$ and the MT-GQLRT$_{\text{sub}}$ have the same ACL for classification as the standard GQLRT. Also note that the ACL of the MT-GQLRT$_{\text{opt}}$ and the MT-GQLRT$_{\text{sub}}$ due to tunning of the width parameter of the Gaussian MT-function (which is performed via simple line search) is linear in the sample size, dimension and the number of grid points taken over $\Omega$.

Finally, an analysis of the compared classifiers for small sample size is provided in [36, Sec. F].

TABLE II

*Signal classification:* ASYMPTOTIC COMPUTATIONAL COMPLEXITY (FLOPS). NOTATION: $p$ IS THE DIMENSION OF THE OBSERVATION VECTORS. $N$ DENOTES THE SAMPLE SIZE. $N_0$ AND $N_1$ ARE THE SIZES OF THE TRAINING SEQUENCES FROM $H_0$ AND $H_1$, RESPECTIVELY. $K_\Omega$ DENOTES THE NUMBER OF GRID POINTS OF THE $\Omega$-AXIS. $I$ DENOTES NUMBER OF ITERATIONS. $K_C$ DENOTES THE NUMBER GRID POINTS OVER WHICH THE TUNING PARAMETER $c$ IN THE TUKEY-GQLRT$_{\text{OPT}}$ CLASSIFIER WAS OPTIMIZED. $M$ DENOTES THE NUMBER OF TRIALS OF THE MONTE-CARLO SIMULATION FOR ESTIMATING THE OPTIMAL THRESHOLD.

| Method | Parameter tuning | Threshold calculation | Classification |
|---|---|---|---|
| MT-GQLRT$_{\text{opt}}$ | $O((N_0+N_1)pK_\Omega)$ | $O(1)$ | $O(Np)$ |
| MT-GQLRT$_{\text{sub}}$ | $O(NpK_\Omega)$ | $O(MNp)$ | $O(Np)$ |
| GQLRT | – | $O(MNp)$ | $O(Np)$ |
| Tukey-GQLRT$_{\text{sub}}$ | – | $O(MNpI)$ | $O(NpI)$ |
| Tukey-GQLRT$_{\text{opt}}$ | $O(((N_0+N_1)p^2+p^3)K_C)$ | $O(1)$ | $O(NpI)$ |
| Median-GQLRT | – | $O(MNp)$ | $O(Np)$ |
| GGD-QLRT | $O((N_0+N_1)pI)$ | $O(MNp)$ | $O(Np)$ |
| SVM | $O((N_0+N_1)pI)$ | $O(MNp)$ | $O(Np)$ |

## VII. Conclusion

In this paper a new test, called MT-GQLRT, for non-Bayesian binary hypothesis testing was developed that applies GQLRT after transformation of the probability distribution of the data. A Bayesian extension of this test was also developed by applying the transformation to the conditional probability distribution of the data. By specifying the MT-function in the Gaussian family of functions the non-Bayesian and Bayesian MT-GQLRTs were successfully applied to robust signal detection and classification, respectively.

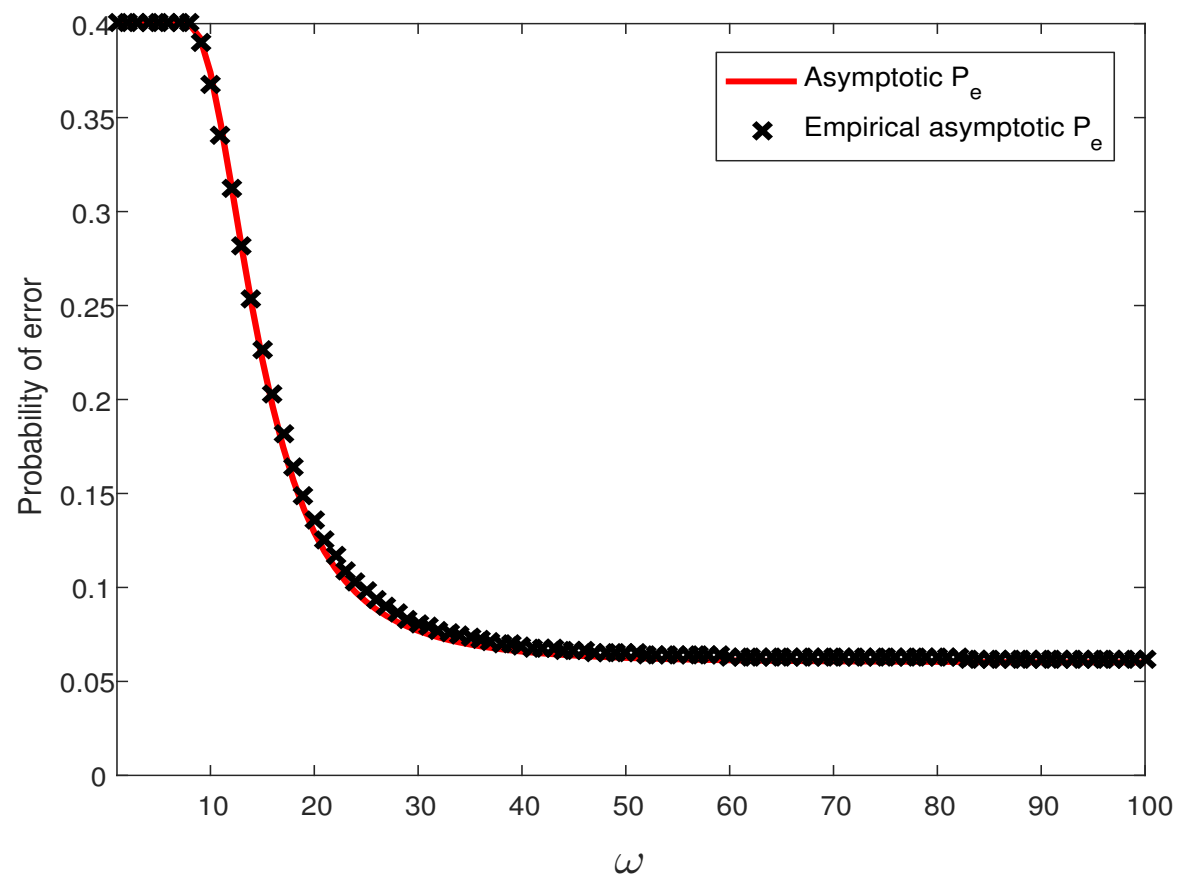

(a)

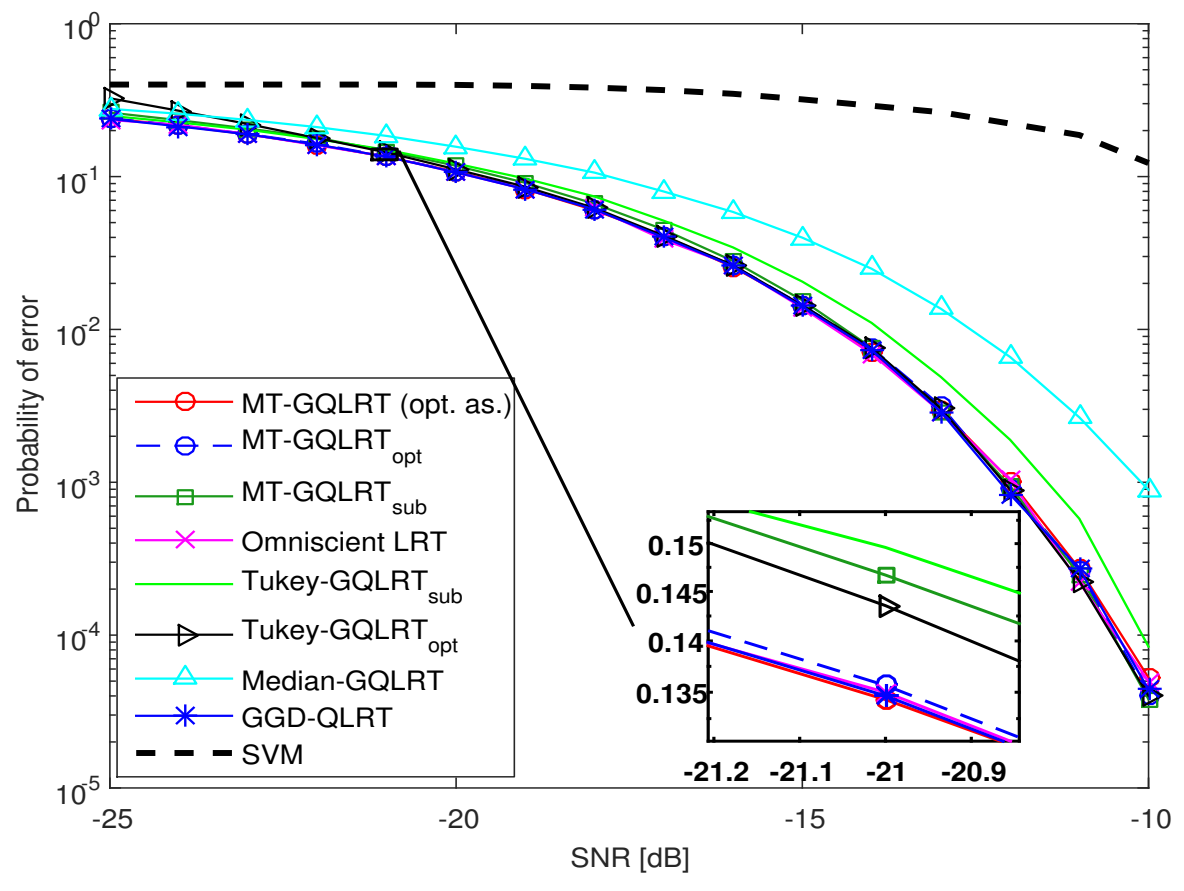

(b)

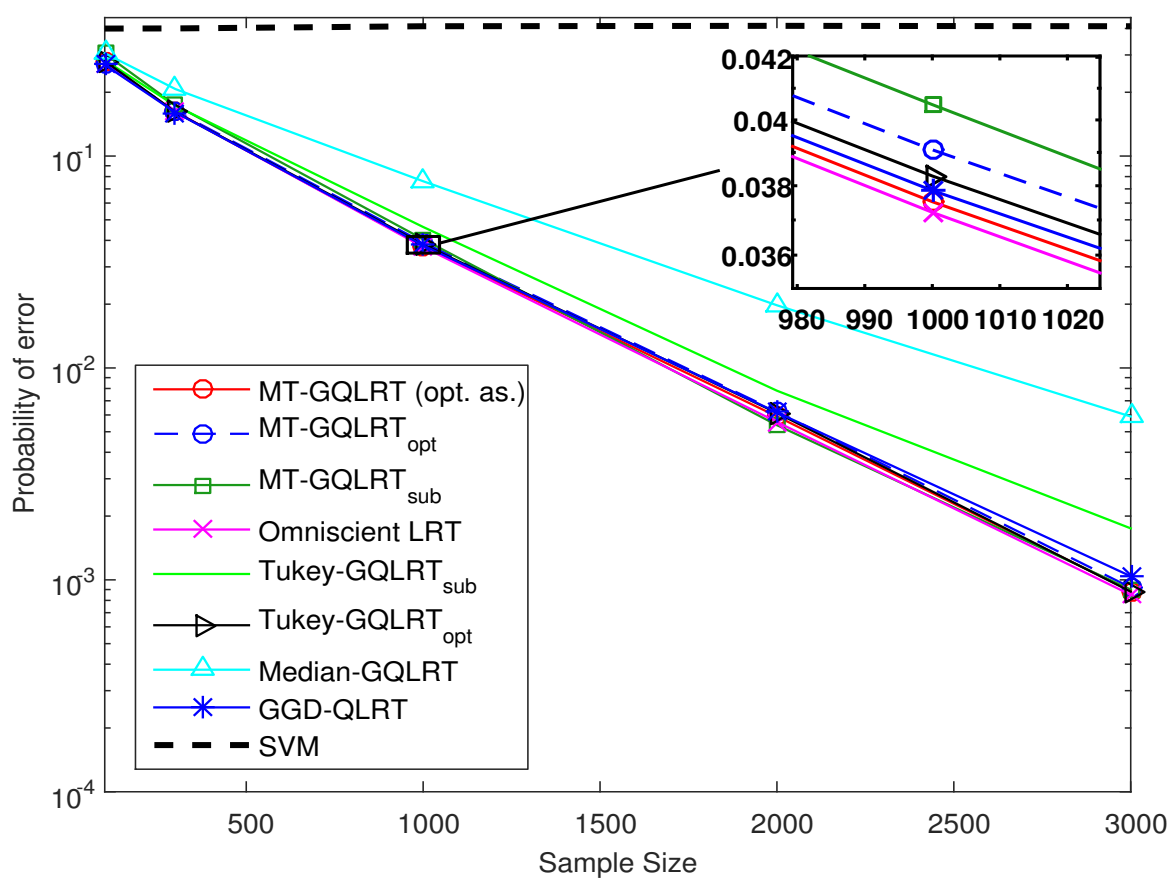

(c)

Fig. 3. **Signal classification in Gaussian noise:** (a) Asymptotic probability of error predicted by the theory (48) and its empirical estimate (49) versus the width parameter $\omega$ of the Gaussian MT-function (50). (b) + (c) Optimal asymptotic error probability of the MT-GQLRT, and the empirical error probabilities of the MT-GQLRT$_{\rm opt}$ and the MT-GQLRT$_{\rm sub}$ as a function of (b) SNR and (c) sample size as compared to the empirical error probabilities of the Tukey-GQLRT$_{\rm sub}$, Tukey-GQLRT$_{\rm opt}$, Median-GQLRT, GGD-QLRT, SVM and the omniscient LRT.

# APPENDIX

In this Appendix, we provide proofs for theorems, propositions and claims that are stated throughout the paper. Furthermore, training-sequences-free procedures for selection of the parameters of the MT-functions (42) and (50) are developed.

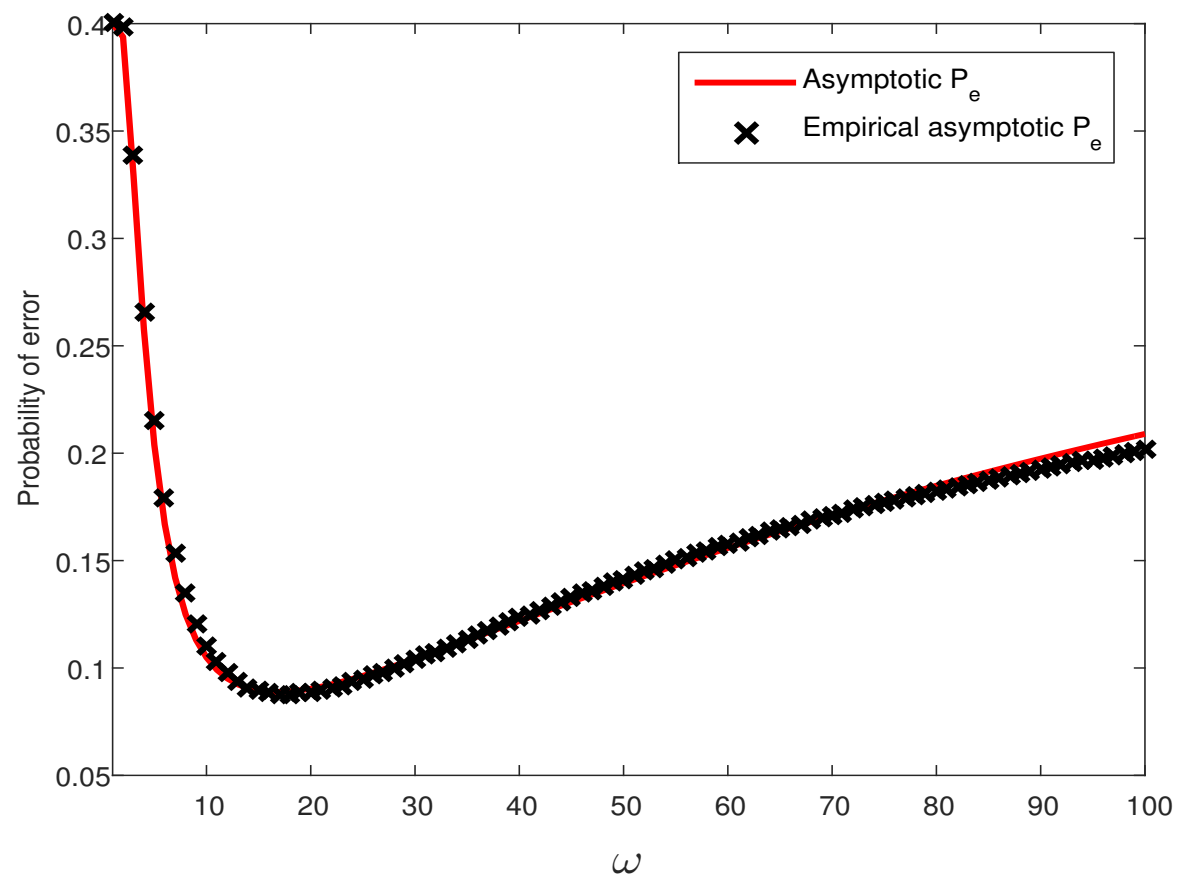

(a)

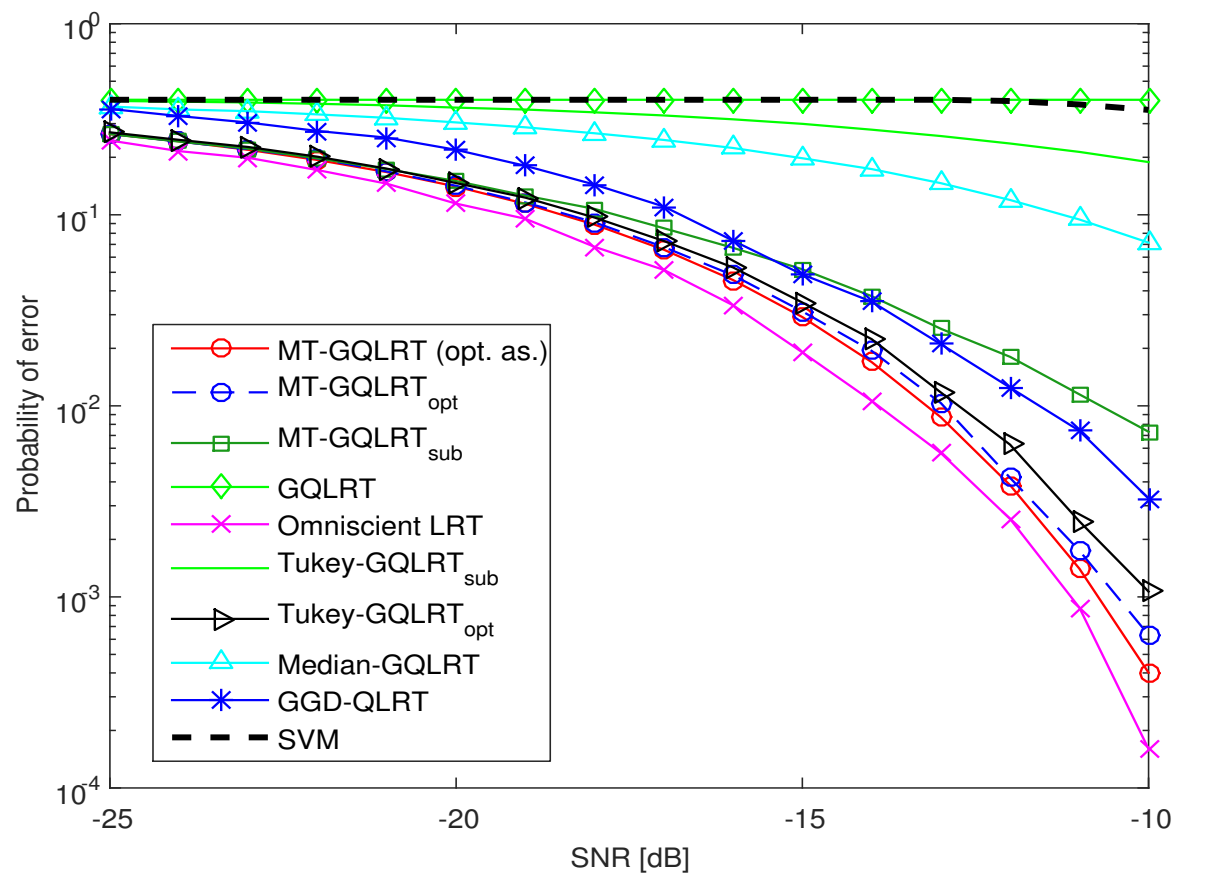

(b)

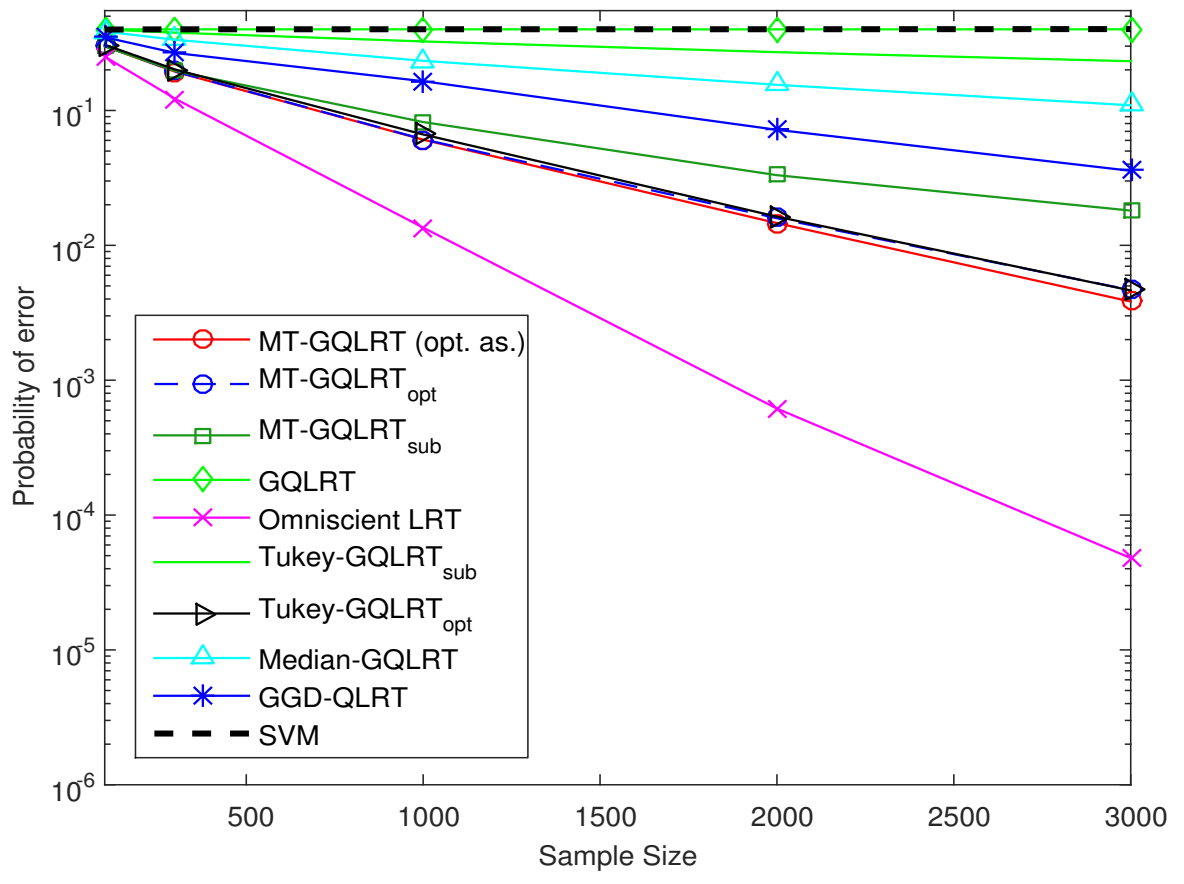

(c)

Fig. 4. **Signal classification in non-Gaussian noise:** (a) Asymptotic probability of error predicted by the theory (48) and its empirical estimate (49) versus the width parameter $\omega$ of the Gaussian MT-function (50). (b) + (c) Optimum asymptotic error probability of the MT-GQLRT, and the empirical error probabilities of the MT-GQLRT$_{\text{opt}}$ and the MT-GQLRT$_{\text{sub}}$ as a function of (b) SNR and (c) sample size as compared to the empirical error probabilities of the GQLRT, Tukey-GQLRT$_{\text{sub}}$, Tukey-GQLRT$_{\text{opt}}$, Median-GQLRT, GGD-QLRT, SVM and the omniscient LRT.

## A. An auxiliary Lemma:

**Lemma 1.** *Assume that* $\mathbf{\Sigma}_k^{(u)}$, $k = 0, 1$ *are non-singular,* $\mathrm{E}[u^2(\mathbf{X}); P_k]$ *and* $\mathrm{E}[\|\mathbf{X}\|^4 u^2(\mathbf{X}); P_k]$ *are finite for* $k = 0, 1$*. Then, the expectations* $A \triangleq \mathrm{E}[u(\mathbf{X})|\psi_u(\mathbf{X})|; P_k]$ *and* $B \triangleq \mathrm{E}[u^2(\mathbf{X})(\psi_u(\mathbf{X}) - \eta_k^{(u)})^2; P_k]$ *are finite for* $k = 0, 1$.

*Proof.* By (17), (19), the non-singularity of $\mathbf{\Sigma}_0^{(u)}$ and $\mathbf{\Sigma}_1^{(u)}$, inequality (1) in [44], and the triangle

inequality:

$$
\begin{aligned}
|\psi_u(\mathbf{X})| &\leq d+\left\|\mathbf{\Sigma}_0^{(u)-1/2}\left(\mathbf{X}-\boldsymbol{\mu}_0^{(u)}\right)\right\|^2+\left\|\mathbf{\Sigma}_1^{(u)-1/2}\left(\mathbf{X}-\boldsymbol{\mu}_1^{(u)}\right)\right\|^2 \\
&\leq d+\left\|\mathbf{\Sigma}_0^{(u)-1/2}\right\|_S^2\left\|\mathbf{X}-\boldsymbol{\mu}_0^{(u)}\right\|^2+\left\|\mathbf{\Sigma}_1^{(u)-1/2}\right\|_S^2\left\|\mathbf{X}-\boldsymbol{\mu}_1^{(u)}\right\|^2 \\
&= d+\lambda_{\min}^{-1}(\mathbf{\Sigma}_0^{(u)})\left\|\mathbf{X}-\boldsymbol{\mu}_0^{(u)}\right\|^2+\lambda_{\min}^{-1}(\mathbf{\Sigma}_1^{(u)})\left\|\mathbf{X}-\boldsymbol{\mu}_1^{(u)}\right\|^2 \\
&\leq d+\lambda_{\min}^{-1}(\mathbf{\Sigma}_0^{(u)})(\|\mathbf{X}\|+\|\boldsymbol{\mu}_0^{(u)}\|)^2+\lambda_{\min}^{-1}(\mathbf{\Sigma}_1^{(u)})(\|\mathbf{X}\|+\|\boldsymbol{\mu}_1^{(u)}\|)^2 \\
&= c_1\|\mathbf{X}\|^2+2c_2\|\mathbf{X}\|+c_3 \triangleq \xi_u(\mathbf{X}), \qquad (51)
\end{aligned}
$$

where $d \triangleq |\log(\det \mathbf{\Sigma}_0^{(u)}/\det \mathbf{\Sigma}_1^{(u)})|$, $\|\cdot\|_S$ denote the spectral norm, $\lambda_{\min}(\cdot)$ denote the minimal eigenvalue of a matrix and $c_1 \triangleq \lambda_{\min}^{-1}(\mathbf{\Sigma}_0^{(u)})+\lambda_{\min}^{-1}(\mathbf{\Sigma}_1^{(u)})$, $c_2 \triangleq \|\boldsymbol{\mu}_0^{(u)}\|\lambda_{\min}^{-1}(\mathbf{\Sigma}_0^{(u)})+\|\boldsymbol{\mu}_1^{(u)}\|\lambda_{\min}^{-1}(\mathbf{\Sigma}_1^{(u)})$, $c_3 \triangleq \|\boldsymbol{\mu}_0^{(u)}\|^2\lambda_{\min}^{-1}(\mathbf{\Sigma}_0^{(u)})+\|\boldsymbol{\mu}_1^{(u)}\|^2\lambda_{\min}^{-1}(\mathbf{\Sigma}_1^{(u)})+d$.

By Definition 1, Hölder's inequality [27] and the assumption that $\mathrm{E}[u^2(\mathbf{X});P_k]$ and $\mathrm{E}[\|\mathbf{X}\|^4u^2(\mathbf{X});P_k]$ are finite for $k=0,1$:

$$
\begin{aligned}
&\mathrm{E}[u(\mathbf{X});P_k]<\infty && (52a)\\
&\mathrm{E}[u^2(\mathbf{X})\|\mathbf{X}\|^2;P_k]\leq\sqrt{\mathrm{E}\left[u^2(\mathbf{X})\|\mathbf{X}\|^4;P_k\right]\mathrm{E}[u^2(\mathbf{X});P_k]}<\infty && (52b)\\
&\mathrm{E}[u^2(\mathbf{X})\|\mathbf{X}\|^3;P_k]\leq\sqrt{\mathrm{E}[u^2(\mathbf{X})\|\mathbf{X}\|^2;P_k]\mathrm{E}[u^2(\mathbf{X})\|\mathbf{X}\|^4;P_k]}<\infty && (52c)\\
&\mathrm{E}[u^2(\mathbf{X})\|\mathbf{X}\|;P_k]\leq\sqrt{\mathrm{E}[u^2(\mathbf{X});P_k]\mathrm{E}[u^2(\mathbf{X})\|\mathbf{X}\|^2;P_k]}<\infty && (52d)
\end{aligned}
$$

for $k=0,1$. Therefore, by (51), (52) and Hölder's inequality:

$$
\begin{aligned}
A &\triangleq \mathrm{E}\left[u(\mathbf{X})|\psi_u(\mathbf{X})|;P_k\right] \qquad (53)\\
&\leq \mathrm{E}\left[u(\mathbf{X})(c_1\|\mathbf{X}\|^2+2c_2\|\mathbf{X}\|+c_3);P_k\right]\\
&\leq \sqrt{\mathrm{E}[u^2(\mathbf{X})(c_1\|\mathbf{X}\|^2+2c_2\|\mathbf{X}\|+c_3)^2;P_k]}<\infty
\end{aligned}
$$

for $k=0,1$. According to (6), (8), (21) and (53) it follows that $|\eta_k^{(u)}|$, $k\in\{0,1\}$ is finite since $|\eta_k^{(u)}|\leq\frac{A}{\mathrm{E}[u(\mathbf{X});P_k]}<\infty$. Moreover, by (51)

$$
\begin{aligned}
(\psi_u(\mathbf{X})-\eta_k^{(u)})^2 &\leq \psi_u^2(\mathbf{X})+2|\psi_u(\mathbf{X})\eta_k^{(u)}|+(\eta_k^{(u)})^2 \qquad (54)\\
&\leq \xi_u^2(\mathbf{X})+2|\eta_k^{(u)}|\xi_u(\mathbf{X})+(\eta_k^{(u)})^2\\
&= \tilde{c}_4\|\mathbf{X}\|^4+\tilde{c}_3\|\mathbf{X}\|^3+\tilde{c}_2\|\mathbf{X}\|^2+\tilde{c}_1\|\mathbf{X}\|+\tilde{c}_0
\end{aligned}
$$

for $k=0,1$, where $\tilde{c}_4 \triangleq c_1^2$, $\tilde{c}_3 \triangleq 2c_1c_2$, $\tilde{c}_2 \triangleq 4c_2^2 + 2|\eta_k^{(u)}|c_1 + 2c_1c_3$, $\tilde{c}_1 \triangleq 2c_2c_3 + 4|\eta_k^{(u)}|c_2$, and $\tilde{c}_0 \triangleq {c_3}^2 + \left(\eta_k^{(u)}\right)^2 + 2|\eta_k^{(u)}|c_3$. Finally, by (52), (54) and the fact that $|\eta_k^{(u)}|$ is finite we conclude that:

$$B \leq \mathrm{E}\left[u^2\left(\mathbf{X}\right)\left(\tilde{c}_4\|\mathbf{X}\|^4 + \tilde{c}_3\|\mathbf{X}\|^3 + \tilde{c}_2\|\mathbf{X}\|^2 + \tilde{c}_1\left\|\mathbf{X}\right\| + \tilde{c}_0\right); P_k\right] < \infty$$

for $k=0,1$. $\square$

### B. *Proof of Theorem 1:*

By (14), (17), (18), (19) and assumption A-1, the test statistic is a non-degenerate random variable that can be written as:

$$T_u = \frac{\frac{1}{N}\sum_{n=1}^{N} u\left(\mathbf{X}_n\right)\psi_u\left(\mathbf{X}_n\right)}{\frac{1}{N}\sum_{n=1}^{N} u\left(\mathbf{X}_n\right)}. \tag{55}$$

Since $\mathbf{X}_n, n = 1, ..., N$ are i.i.d. random vectors and the functions $u(\cdot)$ and $\psi_u(\cdot)$ are real, the products $u(\mathbf{X}_n)\psi_u(\mathbf{X}_n), n = 1, ..., N$ are i.i.d. and real. According to (6), (8) and (21) $u(\mathbf{X})(\psi_u\left(\mathbf{X}\right) - \eta_k^{(u)})$ is a zero-mean random variable under $P_k$ for $k=0,1$. Furthermore, by assumptions A-2, A-3 and Lemma 1 stated in Appendix A, its variance under $P_k$ is finite for any $k=0,1$. Therefore, by the central limit theorem [45] we conclude that the translated and scaled version of the numerator in (55) satisfies:

$$\sqrt{\frac{N}{\tilde{\lambda}_k^{(u)}}}\frac{1}{N}\sum_{n=1}^{N} u\left(\mathbf{X}_n\right)\left(\psi_u\left(\mathbf{X}_n\right) - \eta_k^{(u)}\right) \xrightarrow[N\to\infty]{D} \mathcal{N}\left(0,1\right) \tag{56}$$

$\forall k \in \{0,1\}$, where

$$\tilde{\lambda}_k^{(u)} \triangleq \mathrm{E}\left[u^2\left(\mathbf{X}\right)\left(\psi_u(\mathbf{X}) - \eta_k^{(u)}\right)^2; P_k\right]. \tag{57}$$

Since by Definition 1 $u\left(\mathbf{X}\right)$ is non-negative and $0 < \mathrm{E}\left[u\left(\mathbf{X}\right); P_k\right] < \infty$ for $k=0,1$, by Khinchine's strong law of large numbers [25] we have that the denominator in (55) satisfies:

$$\frac{1}{N}\sum_{n=1}^{N} u\left(\mathbf{X}_n\right) \xrightarrow[N\to\infty]{w.p.1} \mathrm{E}\left[u\left(\mathbf{X}\right); P_k\right] \qquad \forall k \in \{0,1\}. \tag{58}$$

Notice that by Eqs. (8), (22) and (57), $\lambda_k^{(u)} = \tilde{\lambda}_k^{(u)}/(N\mathrm{E}^2\left[u\left(\mathbf{X}\right); P_k\right])$. Therefore, by (55)-(58) and Slutsky's theorem [27]:

$$\frac{T_u - \eta_k^{(u)}}{\sqrt{\lambda_k^{(u)}}} = \frac{\sqrt{\frac{N}{\tilde{\lambda}_k^{(u)}}}\frac{1}{N}\sum_{n=1}^{N} u\left(\mathbf{X}_n\right)\left(\psi_u\left(\mathbf{X}_n\right) - \eta_k^{(u)}\right)}{\left(\frac{1}{N}\sum_{n=1}^{N} u\left(\mathbf{X}_n\right)\right)/\mathrm{E}\left[u\left(\mathbf{X}\right); P_k\right]} \xrightarrow[N\to\infty]{D} \mathcal{N}\left(0,1\right)$$

$\forall k \in \{0,1\}$. $\square$

### C. Proof of proposition 1:

By (14), (19), (25), (26) and assumptions A-1, A-2, the empirical estimators $\hat{\eta}_k^{(u)}$ and $\hat{\lambda}_k^{(u)}$ are non-degenerate random variables that can be written as:

$$\hat{\eta}_k^{(u)} \triangleq \frac{\frac{1}{N_k}\sum_{n=1}^{N_k} u(\mathbf{X}_n^{(k)})\psi_u(\mathbf{X}_n^{(k)})}{\frac{1}{N_k}\sum_{n=1}^{N_k} u(\mathbf{X}_n^{(k)})} \tag{59}$$

and

$$\hat{\lambda}_k^{(u)} \triangleq \frac{\frac{1}{N_k}\sum\limits_{n=1}^{N_k} u^2(\mathbf{X}_n^{(k)})\left(\psi_u^2(\mathbf{X}_n^{(k)}) - 2\psi_u(\mathbf{X}_n^{(k)})\hat{\eta}_k^{(u)} + (\hat{\eta}_k^{(u)})^2\right)}{N\left(\frac{1}{N_k}\sum_{n=1}^{N_k} u(\mathbf{X}_n^{(k)})\right)^2}, \tag{60}$$

respectively. Since $\{\mathbf{X}_n^{(k)}\}_{n=1}^{N_k}$ are i.i.d and the functions $u(\cdot)$ and $\psi_u(\cdot)$ are real, the products $\{u(\mathbf{X}_n^{(k)})\psi_u(\mathbf{X}_n^{(k)})\}_{n=1}^{N_k}$, $\{u^2(\mathbf{X}_n^{(k)})\psi_u^2(\mathbf{X}_n^{(k)})\}_{n=1}^{N_k}$ and $\{u^2(\mathbf{X}_n^{(k)})\psi_u(\mathbf{X}_n^{(k)})\}_{n=1}^{N_k}$ define real i.i.d. sequences.

Furthermore, by Hölder's inequality [27], assumptions A-2, A-3 and Lemma 1 stated in Appendix A we have that the expectations $\mathrm{E}[u(\mathbf{X});P_k]$, $\mathrm{E}[u^2(\mathbf{X});P_k]$, $\mathrm{E}[u(\mathbf{X})|\psi_u(\mathbf{X})|;P_k]$, $\mathrm{E}[u^2(\mathbf{X})\psi_u^2(\mathbf{X});P_k]$ and $\mathrm{E}[u^2(\mathbf{X})|\psi_u(\mathbf{X})|;P_k]$ are finite for any $k \in \{0,1\}$. Therefore, by Khinchine's strong law of large numbers [25]:

$$N_k^{-1}\sum_{n=1}^{N_k} u^i(\mathbf{X}_n^{(k)})\psi_u^j(\mathbf{X}_n^{(k)}) \xrightarrow[N_k\to\infty]{\text{w.p.1}} \mathrm{E}[u^i(\mathbf{X})\psi_u^j(\mathbf{X});P_k], \tag{61}$$

for any $(i,j) \in \{(1,0),(2,0),(1,1),(2,1),(2,2)\}$ and any $k \in \{0,1\}$. Hence, by (8), (21), (22), (59)-(61) and Mann-Wald's Theorem [46] we conclude that

$$\hat{\eta}_k^{(u)} \xrightarrow[N_k\to\infty]{\text{w.p.1}} \eta_k^{(u)} \quad \text{and} \quad \hat{\lambda}_k^{(u)} \xrightarrow[N_k\to\infty]{\text{w.p.1}} \lambda_k^{(u)}, \quad k = 0,1. \tag{62}$$

Therefore, by (23), (24), (62), the continuity of the standard normal tail probability $Q(\cdot)$ and Mann-Wald's Theorem [46] we conclude that $\hat{\alpha}_u \xrightarrow[N_0\to\infty]{\text{w.p.1}} \alpha_u$ and $\hat{\beta}_u \xrightarrow[N_1\to\infty]{\text{w.p.1}} \beta_u$. □

### D. Proof of proposition 3:

Similarly to the proof of Proposition 1 stated in Appendix C, one can verify that under conditions A-1 - A-3, (62) holds. for any $k \in \{0,1\}$. Therefore, by (28), (29), (62), the continuity of $Q(\cdot)$ and Mann-Wald's Theorem [46] we conclude that $\hat{R}^{(u)} \xrightarrow{\text{w.p.1}} R^{(u)}$ as $N_0, N_1 \to \infty$. □

*E. Proof of proposition 4:*

One can verify that if assumption C-1 is satisfied and $\hat{\lambda}_1^{(u)} \neq \hat{\lambda}_0^{(u)}$, then the only two stationary points [47] of $\hat{R}^{(u)}(\cdot)$ are given by

$$t_1^* \triangleq \frac{\hat{\lambda}_0^{(u)}\hat{\eta}_1^{(u)} - \hat{\lambda}_1^{(u)}\hat{\eta}_0^{(u)} - \sqrt{\hat{\lambda}_0^{(u)}\hat{\lambda}_1^{(u)}\hat{s}^{(u)}}}{\hat{\lambda}_0^{(u)} - \hat{\lambda}_1^{(u)}}$$

and

$$t_2^* \triangleq \frac{\hat{\lambda}_0^{(u)}\hat{\eta}_1^{(u)} - \hat{\lambda}_1^{(u)}\hat{\eta}_0^{(u)} + \sqrt{\hat{\lambda}_0^{(u)}\hat{\lambda}_1^{(u)}\hat{s}^{(u)}}}{\hat{\lambda}_0^{(u)} - \hat{\lambda}_1^{(u)}}.$$

Furthermore, $\hat{R}^{(u)}(\cdot)$ is twice differentiable at $t_1^*$ and $t_2^*$, $\frac{d^2\hat{R}^{(u)}}{dt^2}(t_1^*) > 0$ and $\frac{d^2\hat{R}^{(u)}}{dt^2}(t_2^*) < 0$. Hence, by the second derivative test [47], $t_1^*$ is a local minimum of $\hat{R}^{(u)}(\cdot)$, and $t_1^*$ is a local maximum of $\hat{R}^{(u)}(\cdot)$. Therefore, by Fermat's Theorem [47] and the fact that $\hat{R}^{(u)}(\cdot)$ is differentiable at any $t \in \mathbb{R}$, we conclude that exactly one of the following statements is satisfied:

a) $t_1^*$ is a global minimum of $\hat{R}^{(u)}(\cdot)$.

b) $L_{10}\pi_0 = \lim_{t\to-\infty} \hat{R}^{(u)}(t) \leq \hat{R}^{(u)}(r)$ for all $r \in \mathbb{R}$.

c) $L_{01}\pi_1 = \lim_{t\to\infty} \hat{R}^{(u)}(t) \leq \hat{R}^{(u)}(r)$ for all $r \in \mathbb{R}$.

If in addition to assumption C-1, assumption C-2 is satisfied then statement a must hold. Now, if assumption C-2 is not satisfied then statement b or statement c must hold, which means that $\hat{R}^{(u)}(t) > \min(L_{10}\pi_0, L_{01}\pi_1)$ for all $t \in \mathbb{R}$ , i.e. $t_1^*$ is not a global minimum. Furthermore, if assumption C-1 is not satisfied then $\hat{R}^{(u)}(\cdot)$ has no stationary points and, again, $t_1^*$ is not a global minimum. □

*F. Signal detection: Training-sequences-free procedure for selection of the Gaussian MT-function width parameter:*

In the following, a data-driven procedure for selection of the Gaussian MT-function (42) width parameter $\omega$ is developed that does not require training sequences. This procedure, which is based on a weak-signal assumption, has the property that it prevents significant loss in the asymptotic local power sensitivity to change in signal variance, relative to the omniscient LRT, when the observations are normally distributed.

By (38), (39) and (42) one can verify that when $S_n \sim \mathcal{CN}\left(0, \sigma_S^2\right)$ and $\mathbf{W}_n \sim \mathcal{CN}\left(\mathbf{0}, \sigma_{\mathbf{W}}^2\mathbf{I}_p\right)$, the asymptotic powers of the omniscient LRT and the proposed test (37) at a fixed test size $\alpha$ are given by:

$$\beta_{LRT} = Q\left(\frac{Q^{-1}(\alpha)\,\sigma_{\mathbf{W}}^2 - \sqrt{N}\sigma_S^2}{\sigma_S^2 + \sigma_{\mathbf{W}}^2}\right)$$

and

$$\beta_{u_G}(\omega) = Q\left(\frac{Q^{-1}(\alpha)\,\sigma_{\mathbf{W}}^2 - \sqrt{N}\sigma_S^2 G(\omega,\sigma_{\mathbf{W}}^2)}{\sigma_S^2+\sigma_{\mathbf{W}}^2}\right),$$

respectively, where $G(\omega,\sigma_{\mathbf{W}}^2) \triangleq \left(\frac{\sqrt{1+2\sigma_{\mathbf{W}}^2/\omega^2}}{1+\sigma_{\mathbf{W}}^2/\omega^2}\right)^{p-1}$.

The asymptotic local power sensitivity is defined as the gradient of the power w.r.t. the signal variance at $\sigma_S^2=0$. Hence, the relative local power sensitivity is quantified by the ratio:

$$R(\omega,\sigma_{\mathbf{W}}^2) \triangleq \frac{\partial\beta_{u_G}}{\partial\sigma_S^2}\Big/\frac{\partial\beta_{LRT}}{\partial\sigma_S^2}\bigg|_{\sigma_S^2=0} = \frac{\sqrt{N}G(\omega,\sigma_{\mathbf{W}}^2)}{\sqrt{N}+Q^{-1}(\alpha)} + d, \tag{63}$$

where $d \triangleq Q^{-1}(\alpha)/(\sqrt{N}+Q^{-1}(\alpha))$.

Thus, in order to prevent significant loss in (63), we propose to choose the width parameter $\omega$ that solves the equation:

$$R(\omega,\hat{\sigma}_{\mathbf{Y}}^2) = r, \tag{64}$$

where $d << r < 1$ is a predefined constant, $\hat{\sigma}_{\mathbf{Y}}^2 \triangleq \sqrt{\frac{c^2}{p-1}\sum_{k=1}^{p}\hat{\sigma}_k^2}$ is an empirical estimate of the unknown noise variance $\sigma_{\mathbf{W}}^2$, and $\hat{\sigma}_k^2 = \text{MAD}^2(\{\text{Re}([\mathbf{Y}_n]_k)\}_{n=1}^N) + \text{MAD}^2(\{\text{Im}([\mathbf{Y}_n]_k)\}_{n=1}^N)$ is a robust median absolute deviation (MAD) estimate of variance [29]. Here, $\mathbf{Y}_n \triangleq \mathbf{P}_{\mathbf{a}}^{\perp}\mathbf{X}_n$ and $c \triangleq 1/\text{erf}^{-1}(3/4)$ ensures consistency of the variance estimate for normally distributed data [29] under both hypotheses. One can verify that the solution of (64) is given by:

$$\omega_0^2 = \left(1/\sqrt{\zeta}-1\right)\hat{\sigma}_{\mathbf{Y}}^2, \tag{65}$$

where $\zeta \triangleq 1-(r-(1-r)Q^{-1}(\alpha)/\sqrt{N})^{2/(p-1)}$.

In the considered example, the parameter $r$ in (64) was set to 0.9. We note that a similar strategy for selection of $\omega$ was carried out in [48] for a different test that was developed for constant-false-alarm-rate (CFAR) radar target detection in non-spherical noise. Unlike the procedure described above, the selection algorithm in [48] utilizes a training sequence (secondary data) from the null hypothesis.

### *G. Signal classification: Training-sequences-free procedure for selection of the Gaussian MT-function width parameter:*

In the following, a data-driven procedure for selection of the Gaussian MT-function (50) width parameter $\omega$ is developed. This training-sequences-free procedure controls the asymptotic error probability of the proposed test (47) relative to this of the LRT when the observations are normally distributed.

Using (48), one can verify that when $\mathbf{W} \sim \mathcal{CN}(\mathbf{0}, \sigma^2_{\mathbf{W}}\mathbf{I})$, the minimum asymptotic probability of error of the MT-GQLRT (47) w.r.t. the threshold parameter is given by:

$$P_e^{(u_G)}\left(\omega, \sigma^2_{\mathbf{W}}\right) = \sum_{k=0}^{1} \pi_k Q\left(\frac{G_{u_G}(\omega, \sigma^2_{\mathbf{W}})}{2} + \frac{(-1)^k \log \frac{\pi_0}{\pi_1}}{G_{u_G}(\omega, \sigma^2_{\mathbf{W}})}\right), \tag{66}$$

where $G_u(\omega, \sigma^2_{\mathbf{W}}) \triangleq G_L(\sigma^2_{\mathbf{W}})(\omega\sqrt{\omega^2 + \sigma^2_{\mathbf{W}}}/(\omega^2 + 2\sigma^2_{\mathbf{W}}))^{p-2}$ and $G_L(\sigma^2_{\mathbf{W}}) \triangleq \sqrt{N}\|\mathbf{a}_0 - \mathbf{a}_1\|/\sqrt{2\sigma^2_{\mathbf{W}}}$. When the observations are normally distributed, the LRT coincides with the GQLRT. Therefore, in this case, the LRT is obtained from the MT-GQLRT for $u(\mathbf{x}) = 1$. Notice that the Gaussian MT-function (50) satisfies $u_G(\mathbf{x}; \omega) \to 1$ as $\omega \to \infty$. Hence, we conclude that, for Gaussian observations, the corresponding error probability of the LRT, denoted as $P_e^{(LRT)}(\sigma^2_{\mathbf{W}})$, can be obtained from (66) by taking the limit of $P_e^{(u_G)}(\omega, \sigma^2_{\mathbf{W}})$ as $\omega$ goes to infinity. A closed from expression is easily obtained by replacing $G_u(\omega, \sigma^2_{\mathbf{W}})$ in (66) with $G_L(\sigma^2_{\mathbf{W}})$.

The asymptotic relative error probability is defined here as:

$$R(\omega, \sigma^2_{\mathbf{W}}) \triangleq P_e^{(\mathrm{LRT})}(\sigma^2_{\mathbf{W}})/P_e^{(u)}(\omega, \sigma^2_{\mathbf{W}}). \tag{67}$$

Thus, in order to prevent low values of (67), we propose to chose $\omega$ that solves the equation $R(\omega, \sigma^2_{\mathbf{W}}) = r$ where $0 << r < 1$ is some predefined constant.

In practice the noise variance $\sigma^2_{\mathbf{W}}$ is unknown. Hence, we propose to estimate this quantity in the following manner. Similarly to (46), one can verify using (8), (11), (43) and (50) that under the assumption of Gaussian noise, the MT-covariances under both hypotheses are given by:

$$\mathbf{\Sigma}_k^{(u_G)}(\omega) = \frac{\sigma^4_{\mathbf{W}}}{\omega^2 + \sigma^2_{\mathbf{W}}}\mathbf{P}_{\mathbf{A}} + \frac{\sigma^2_{\mathbf{W}}\omega^2}{\omega^2 + \sigma^2_{\mathbf{W}}}\mathbf{I}, \tag{68}$$

$k = 0, 1$. A robust estimator of $\sigma^2_{\mathbf{W}}$ can be constructed by taking trace on both sides of (68), extracting $\sigma^2_{\mathbf{W}}$, and replacing the MT-covariance $\mathbf{\Sigma}_k^{(u_G)}(\omega)$ by its empirical estimate $\hat{\mathbf{\Sigma}}^{(u_G)}(\omega)$, which is obtained according to (13) using the same sequence of samples that comprises the test-statistic in (47). This results in the following estimate of noise variance:

$$\hat{\sigma}^2_{\mathbf{W}}(\omega) \triangleq \frac{d(\omega) - p\omega^2 + \sqrt{(d(\omega) - p\omega^2)^2 + 8\omega^2 d(\omega)}}{4}, \tag{69}$$

where $d(\omega) \triangleq \mathrm{trace}\{\hat{\mathbf{\Sigma}}_k^{(u_G)}(\omega)\}$. Thus, the desired parameter $\omega_0$ is obtained by solving the equation $R(\omega, \hat{\sigma}^2_{\mathbf{W}}(\omega)) = r$. In the considered example, this equation was solved numerically via search over $K_\Omega = 100$ grid points of the interval $\Omega = [1, 100]$, where the asymptotic relative error parameter was set to $r = 0.9$.

# Binary Hypothesis Testing via Measure Transformed Quasi Likelihood Ratio Test: Supplementary Material

Nir Halay*, Koby Todros* and Alfred O. Hero†
*Ben-Gurion University of the Negev, †University of Michigan

In this supplementary material document, we provide algebraic manipulations showing the structure of the measure-transformed covariance of the noise component under the detection problem (31). Furthermore, implementation details for some of the compared methods and an additional performance analysis for small sample size are provided. Finally, a modified scale-invariant version of the proposed detector (37) is presented.

### A. Signal detection example: structure of the noise MT-covariance

In this section we provide detailed algebraic manipulation showing the structure of the measure-transformed covariance of the noise component, as stated below Eq. (35). According to (8), (10), (11) and (33) the MT-covariance of the spherically symmetric noise component is given by:

$$\boldsymbol{\Sigma}_{\mathbf{W}}^{(u)} \triangleq \mathrm{E}\left[\mathbf{W}\mathbf{W}^H g\left(\left\|\mathbf{P}_{\mathbf{a}}^{\perp}\mathbf{W}\right\|\right); P_{\mathbf{W}}\right]/c, \tag{S-1}$$

where $P_{\mathbf{W}}$ denote the probability distribution of $\mathbf{W}$ and the constant $c \triangleq \mathrm{E}[g(\|\mathbf{P}_{\mathbf{a}}^{\perp}\mathbf{W}\|); P_{\mathbf{W}}]$. Define the unitary matrix $\mathbf{U} \triangleq [\mathbf{a}, \mathbf{V}_{\mathbf{a}}^{\perp}]$ (recall that $\mathbf{a}$ is unit-norm), where $\mathbf{V}_{\mathbf{a}}^{\perp} \in \mathbb{C}^{p\times p-1}$ is the orthonormal complement of $\mathbf{a}$. Notice that since $\mathbf{W}$ is spherically distributed, it has the same probability distribution as $\mathbf{UW}$, hence:

$$\begin{aligned} A &\triangleq \mathrm{E}\left[\mathbf{W}\mathbf{W}^H g\left(\left\|\mathbf{P}_{\mathbf{a}}^{\perp}\mathbf{W}\right\|\right); P_{\mathbf{W}}\right] \\ &= \mathbf{U}\mathrm{E}\left[\mathbf{W}\mathbf{W}^H g\left(\left\|\mathbf{P}_{\mathbf{a}}^{\perp}\mathbf{U}\mathbf{W}\right\|\right); P_{\mathbf{W}}\right]\mathbf{U}^H \\ &= \mathbf{U}\mathrm{E}\left[\mathbf{W}\mathbf{W}^H g\left(\left\|\left[\mathbf{0}, \mathbf{V}_{\mathbf{a}}^{\perp}\right]\mathbf{W}\right\|\right); P_{\mathbf{W}}\right]\mathbf{U}^H. \end{aligned} \tag{S-2}$$

The third equality in (S-2) follows directly from the property that $\mathbf{P}_{\mathbf{a}}^{\perp} = \mathbf{V}_{\mathbf{a}}^{\perp}\mathbf{V}_{\mathbf{a}}^{\perp H}$. Now, let $\mathbf{W} = [W_1, W_2, \ldots, W_p]^T$, and define $\mathbf{W}_2 \triangleq [W_2, \ldots, W_p]^T$. Since $\mathbf{V}_{\mathbf{a}}^{\perp H}\mathbf{V}_{\mathbf{a}}^{\perp} = \mathbf{I}_{p-1}$, we conclude that

$$\begin{aligned} g\left(\left\|\left[0, \mathbf{V}_{\mathbf{a}}^{\perp}\right]\mathbf{W}\right\|\right) &= g\left(\left\|\mathbf{V}_{\mathbf{a}}^{\perp}\mathbf{W}_2\right\|\right) \\ &= g\left(\sqrt{\mathbf{W}_2^H\mathbf{V}_{\mathbf{a}}^{\perp H}\mathbf{V}_{\mathbf{a}}^{\perp}\mathbf{W}_2}\right) = g(\|\mathbf{W}_2\|). \end{aligned} \tag{S-3}$$

Hence, by (S-2) and (S-3) we obtain that

$$A = \mathbf{U}\begin{bmatrix} \mathrm{E}\left[|W_1|^2 g(\|\mathbf{W}_2\|); P_{\mathbf{W}}\right] & \mathrm{E}\left[W_1\mathbf{W}_2^H g(\|\mathbf{W}_2\|); P_{\mathbf{W}}\right] \\ \mathrm{E}[W_1^*\mathbf{W}_2 g(\|\mathbf{W}_2\|); P_{\mathbf{W}}] & \mathrm{E}\left[\mathbf{W}_2\mathbf{W}_2^H g(\|\mathbf{W}_2\|); P_{\mathbf{W}_2}\right] \end{bmatrix}\mathbf{U}^H. \tag{S-4}$$

Since $\mathbf{W}$ is spherically symmetric, it has the same distribution as $\tilde{\mathbf{W}} \triangleq [W_1, -W_2, ..., -W_p]^T$. Therefore, we conclude that:

$$\mathrm{E}\left[W_1\mathbf{W}_2^H g\left(\|\mathbf{W}_2\|\right); P_{\mathbf{W}}\right] = \mathbf{0}^T, \tag{S-5}$$

Furthermore, since any sub-vector of a spherically symmetric random vector is also spherically symmetric [s1], then $\mathbf{W}_2$ defined above must be spherically symmetric, and therefore, $\mathrm{E}[\mathbf{W}_2\mathbf{W}_2^H g(\|\mathbf{W}_2\|); P_{\mathbf{W}_2}] = \mathbf{Q}\mathrm{E}[\mathbf{W}_2\mathbf{W}_2^H g(\|\mathbf{W}_2\|); P_{\mathbf{W}_2}]\mathbf{Q}^H$ for any unitary matrix $\mathbf{Q} \in \mathbb{C}^{p-1\times p-1}$. Hence,

$$\mathrm{E}\left[\mathbf{W}_2\mathbf{W}_2^H g(\|\mathbf{W}_2\|); P_{\mathbf{W}_2}\right] = \frac{1}{p-1}\mathrm{E}\left[\|\mathbf{W}_2\|^2 g(\|\mathbf{W}_2\|); P_{\mathbf{W}_2}\right]\mathbf{I}_{p-1}. \tag{S-6}$$

Finally, by the definition of $\mathbf{U}$, (S-1), (S-4)-(S-6) and the fact that $\mathbf{P}_{\mathbf{a}}^{\perp} = \mathbf{V}_{\mathbf{a}}^{\perp}\mathbf{V}_{\mathbf{a}}^{\perp H}$ we obtain

$$\begin{aligned} \boldsymbol{\Sigma}_{\mathbf{W}}^{(u)} &= \mathrm{E}\left[\mathbf{W}\mathbf{W}^H g\left(\left\|\mathbf{P}_{\mathbf{a}}^{\perp}\mathbf{W}\right\|\right); P_{\mathbf{W}}\right]/c \\ &= \mathrm{E}\left[|W_1|^2 g\left(\|\mathbf{W}_2\|\right); P_{\mathbf{W}}\right]\mathbf{a}\mathbf{a}^H/c \\ &\quad + \mathrm{E}\left[\|\mathbf{W}_2\|^2 g\left(\|\mathbf{W}_2\|\right); P_{\mathbf{W}}\right]\mathbf{P}_{\mathbf{a}}^{\perp}/\left(c(p-1)\right) \\ &= r_0^{(u)}\mathbf{a}\mathbf{a}^H + r_1^{(u)}\mathbf{I}_p, \end{aligned} \tag{S-7}$$

where $r_0^{(u)} \triangleq \mathrm{E}[(|W_1|^2 - \frac{\|\mathbf{W}_2\|^2}{p-1})g(\|\mathbf{W}_2\|); P_{\mathbf{W}}]/c$ and $r_1^{(u)} \triangleq \frac{1}{p-1}\mathrm{E}[\|\mathbf{W}_2\|^2 g(\|\mathbf{W}_2\|); P_{\mathbf{W}}]/c$.

### B. Maximum likelihood estimation of GGD parameters - Signal detection example

In this section we provide exact implementation details of the ML estimator for the GGD parameters comprising density estimation based detector in Section VI-A. We consider the family of generalized Gaussian distributions (GGD) [s1], with densities:

$$\tilde{f}\left(\mathbf{X}; \boldsymbol{\theta}_k\right) = \frac{s_k\Gamma(p)}{\pi^p\Gamma(p/s_k)}\frac{1}{|\boldsymbol{\Sigma}_k|}\exp\left(-\left(\mathbf{x}^H\boldsymbol{\Sigma}_k^{-1}\mathbf{x}\right)^{s_k}\right),$$

where $k = 0, 1$ denotes a hypothesis index, $\Gamma(\cdot)$ is the gamma function, $\boldsymbol{\Sigma}_0 \triangleq \gamma\mathbf{I}$, $\boldsymbol{\Sigma}_1 \triangleq \alpha\mathbf{a}\mathbf{a}^H + \beta\mathbf{I}$, $\boldsymbol{\theta}_0 \triangleq [\gamma, s_0]^T$ and $\boldsymbol{\theta}_1 \triangleq [\alpha, \beta, s_1]^T$. By equating the gradient of the objective functions $J_k\left(\boldsymbol{\theta}_k\right) \triangleq \sum_{n=1}^{N_k}\log\tilde{f}(\mathbf{X}_n^{(k)}; \boldsymbol{\theta}_k)$, $k = 0, 1$ to zero, one can verify that the ML estimators of the vectors $\boldsymbol{\theta}_0$ and $\boldsymbol{\theta}_1$ are the solutions of the equations

$$\hat{\gamma} = \left(\frac{\hat{s}_0}{pN_0}\sum_{n=1}^{N_0}\|\mathbf{X}_n^{(0)}\|^{2\hat{s}_0}\right)^{1/\hat{s}_0},$$

$$\hat{s}_0 = p\psi_0\left(p/\hat{s}_0\right)\left(\frac{\hat{s}_0}{N_0}\sum_{n=1}^{N_0}\frac{\|\mathbf{X}_n^{(0)}\|^{2\hat{s}_0}}{\hat{\gamma}^{\hat{s}_0}}\log\frac{\|\mathbf{X}_n^{(0)}\|^2}{\hat{\gamma}^{\hat{s}_0}} - 1\right)^{-1}$$

for $k = 0$, and

$$\hat{\alpha} = \frac{\hat{s}_1}{N_1\hat{\beta}^{\hat{s}_1-1}}\sum_{n=1}^{N_1}\nu_n^{\hat{s}_1}\left|\mathbf{a}^H\mathbf{X}_n^{(1)}\right|^2 - \hat{\beta},$$

$$\hat{\beta}=\left[\frac{\hat{s}_1\sum_{n=1}^{N_1}\nu_n^{\hat{s}_1-1}\left(\left\|\mathbf{X}_n^{(1)}\right\|^2-\left[1-\left(\frac{\hat{\beta}}{\hat{\alpha}+\hat{\beta}}\right)^2\right]\left|\mathbf{a}^H\mathbf{X}_n^{(1)}\right|^2\right)}{N_1\left(\hat{\alpha}\left(p-1\right)+p\hat{\beta}\right)/\left(\hat{\alpha}+\hat{\beta}\right)}\right]^{1/\hat{s}_1},$$

$$\hat{s}_1=p\psi_0\left(p/\hat{s}_1\right)\left[\frac{1}{N_1}\sum_{n=1}^{N_1}\frac{\nu_n^{\hat{s}_1}}{\hat{\beta}^{\hat{s}_1}}\log\left(\frac{\nu_n^{\hat{s}_1}}{\hat{\beta}^{\hat{s}_1}}\right)-1\right]^{-1}$$

for $k=1$, where $\psi_0(\cdot)$ is the polygamma function of order 0 and $\nu_n\triangleq\|\mathbf{X}_n^{(1)}\|^2-\hat{\alpha}|\mathbf{a}^H\mathbf{X}_n^{(1)}|^2/(\hat{\alpha}+\hat{\beta})$. The solution of these equations was obtained by fixed-point iteration. The maximum number of iterations and the stopping criterion were set to 1000 and $|J_k(\boldsymbol{\theta}_k^{(l)})-J_k(\boldsymbol{\theta}_k^{(l-1)})|/|J_k(\boldsymbol{\theta}_k^{(l-1)})|<10^{-6}$, respectively, where $\boldsymbol{\theta}_k^{(l)}$ denotes the estimates of $\boldsymbol{\theta}_k$ at iteration index $l$. The initial conditions of $\hat{s}_0$, $\hat{s}_1$, $\hat{\alpha}$ and $\hat{\beta}$ were set to 1, 1, 0 and $\hat{\gamma}$, respectively.

### C. Implementation of the SVM based detector

In this section we provide implementation details of the SVM based detector we compared to in the signal detection example in Subsection VI-A. First, a radial basis function (RBF) kernel SVM was trained using MATLAB function "*fitcsvm.m*". We used the same two training sequences utilized by the proposed MT-GQLRT. Notice that according to (31) we are interested in testing between $H_0$ and $H_1$ based on a sequence of samples (observations) $\mathbf{X}_1,\ldots,\mathbf{X}_N$ from $P$. In other words, we need to decide whether the whole set of observations is associated with $H_0$ or $H_1$. Now, the SVM method is inherently designed to classify each observation vector separately. Therefore, after the training stage, the SVM based detector was implemented in two steps. First, each observation vector $\mathbf{X}_n\in\{\mathbf{X}_1,\ldots,\mathbf{X}_N\}$ was classified into $H_0$ or $H_1$ using SVM. This step was performed using MATLAB function "*predict.m*". Second, the final decision was carried out by comparing the averaged SVM output to a threshold value. More specifically, let $C_{\text{SVM}}(\mathbf{X})$ denote the SVM output, such that $C_{\text{SVM}}(\mathbf{X})=0$ when $\mathbf{X}$ is classified into $H_0$ and $C_{\text{SVM}}(\mathbf{X})=1$, otherwise. The final SVM based decision rule is defined as:

$$T_{\text{SVM}}(\mathbf{X}_1,\ldots,\mathbf{X}_N)\triangleq\frac{1}{N}\sum_{n=1}^{N}C_{\text{SVM}}(\mathbf{X}_n)\underset{H_0}{\overset{H_1}{\gtrless}}t, \tag{S-8}$$

where $t\in[0,1]$ denotes a threshold. Notice that when $t=0.5$, we obtain a majority voting based rule, i.e., the alternative hypothesis $H_1$ is accepted if the majority (more than 50%) of the observations are classified by the SVM into $H_1$. The threshold value $t$ was determined using Monte-Carlo simulations to satisfy a fixed false alarm rate.

### D. Implementation of Tukey's bi-square M-estimator of location - Signal classification example:

This section provides exact implementation details for the choice of the tuning parameter in Tukey's bi-square M-estimator of scatter. This estimator comprises the test statistic of the robust GQLRT extension in Subsection VI-B. The considered Tukey's bi-square M-estimator of location minimizes the following objective function

$$J_\rho\left(\mathbf{a}\right)\triangleq\sum_{n=1}^{N}\rho\left(\|\mathbf{X}_n-\mathbf{a}\|\right),$$

where $\rho\left(r\right)\triangleq1-\left(1-\left(\frac{r}{c}\right)^2\right)^3\mathbb{1}_{[0,c]}\left(|r|\right)$ is Tukey's bi-square loss function, $c$ is a tuning constant, and $\mathbb{1}_A\left(\cdot\right)$ denotes the indicator function of a set $A$. By equating the gradient of the objective function $J_\rho\left(\mathbf{a}\right)$ to zero, Tukey's bi-square M-estimator of location is the solution of the equation $\hat{\mathbf{a}}=\sum_{n=1}^{N}\mathbf{X}_n w(\mathbf{X}_n,\hat{\mathbf{a}})/\sum_{n=1}^{N}w(\mathbf{X}_n,\hat{\mathbf{a}})$, obtained by fixed-point iteration, where the weight function $w\left(\mathbf{X}_n,\mathbf{a}\right)\triangleq(1-\frac{\|\mathbf{X}_n-\mathbf{a}\|^2}{c^2})^2\mathbb{1}_{[0,c]}(\|\mathbf{X}_n-\mathbf{a}\|)$. Here, the fixed-point iteration was initialized by the robust median estimator of location. The maximum number of iterations and the stopping criterion in the fixed-point iteration were set to 100 and $\|\hat{\mathbf{a}}_l-\hat{\mathbf{a}}_{l-1}\|\,/\,\|\hat{\mathbf{a}}_{l-1}\|<10^{-6}$, respectively, where $l$ denotes an iteration index. Notice that unlike Tukey's location estimator, the empirical MT-mean (12) does not involve iterative optimization. The tuning parameter $c$ is selected via two different methods:

*1) Method 1:* We chose the parameter $c$ to ensure fixed asymptotic relative efficiency (ARE) [s2] of the location estimate, relative to the CRLB under nominal Gaussian distribution. Here, we set $c\triangleq\tilde{c}\hat{\sigma}$, where $\hat{\sigma}\triangleq\sqrt{\frac{1}{p}\sum_{k=1}^{p}\hat{\sigma}^2_{X_k}}$ is a robust median absolute deviation (MAD) estimate of variance [s3], $\hat{\sigma}^2_{X_k}=\gamma^2[\text{MAD}^2(\{\text{Re}(\mathbf{X}_{k,n})\}_{n=1}^{N})+\text{MAD}^2(\{\text{Im}(\mathbf{X}_{k,n})\}_{n=1}^{N})]$, and $\gamma\triangleq1/\text{erf}^{-1}\left(3/4\right)$. The constant $\gamma$ ensures consistency of the scale estimate under normally distributed data [s3]. The ARE of the considered Tukey bi-square M-estimator, defined as the ratio between the traces of the CRLB and the asymptotic MSE under Gaussian distribution, is given by:

$$\text{ARE}\left(\tilde{c}\right)=\frac{\text{E}^2\left[\left(1-\left(R/\tilde{c}\right)^2\right)\left(\left(2/p+1\right)\left(R/\tilde{c}\right)^2-1\right)\mathbb{1}_{[0,\tilde{c}]}\left(R\right)\right]}{\text{E}\left[\left(1-\left(R/\tilde{c}\right)^2\right)^4R^2\mathbb{1}_{[0,\tilde{c}]}\left(R\right)/p\right]},$$

where $\sqrt{2}R$ is a chi distributed random variable with $2p$ degrees of freedom. Using this formula, the parameter $\tilde{c}$ was set to achieve ARE of 95% in all simulation examples. For the considered dimension $p=10$ we obtained $\tilde{c}\approx6.2$.

*2) Method 2:* Here, we assume that training sequences $\mathbf{X}_n^{(k)}$, $n=1,\ldots,N_k$, $k=0,1$ from $P_0$ and $P_1$ are available. An optimal choice of the tuning parameter $c$ within some interval $C$ would be the one that minimizes the empirical estimate of the asymptotic probability of error.

One can verify that the test statistic $T_{\text{Tukey}}\triangleq\text{Re}\{(\mathbf{a}_1-\mathbf{a}_0)^H\hat{\mathbf{a}}\}$ under $H_k$, $k=0,1$ is asymptotically normal with mean $\kappa_k\triangleq(\boldsymbol{\theta}_1-\boldsymbol{\theta}_1)^T\boldsymbol{\theta}_k$ and variance $\gamma_k\triangleq(\boldsymbol{\theta}_1-\boldsymbol{\theta}_0)^T\mathbf{F}_k^{-1}\mathbf{E}_k\mathbf{F}_k^{-1}\left(\boldsymbol{\theta}_1-\boldsymbol{\theta}_0\right)/N$, where

$$\boldsymbol{\theta}_k\triangleq[\text{Re}\{\mathbf{a}_k\}^T,\text{Im}\{\mathbf{a}_k\}^T]^T,$$

$$\mathbf{E}_k\triangleq\text{E}\left[d_k^4\left(\mathbf{Y}-\boldsymbol{\theta}_k\right)\left(\mathbf{Y}-\boldsymbol{\theta}_k\right)^T\mathbb{1}_{[0,\infty)}\left(d_k\right);P_k\right],$$

$$\mathbf{F}_k\triangleq\text{E}\left[\left(\frac{4d_k}{c^2}\left(\mathbf{Y}-\boldsymbol{\theta}_k\right)\left(\mathbf{Y}-\boldsymbol{\theta}_k\right)^T-d_k^2\mathbf{I}_{2p}\right)\mathbb{1}_{[0,\infty)}\left(d_k\right);P_k\right],$$

$k=0,1$, $d_k \triangleq 1-\|\mathbf{Y}-\boldsymbol{\theta}_k\|^2/c^2$, and $\mathbf{Y} \triangleq [\text{Re}\{\mathbf{X}\}^T, \text{Im}\{\mathbf{X}\}^T]^T$. Hence, the asymptotic probability of error associated with this method is given by

$$P_e \triangleq \pi_0 Q\left((t-\kappa_0)/\sqrt{\gamma_0}\right) + \pi_1 Q\left((\kappa_1 - t)/\sqrt{\gamma_1}\right),$$

where $t$ denotes a threshold. An empirical estimate of the variance $\gamma_k$, $k=0,1$, is defined as:

$$\hat{\gamma}_k \triangleq (\boldsymbol{\theta}_1 - \boldsymbol{\theta}_0)^T \hat{\mathbf{F}}_k^{-1} \hat{\mathbf{E}}_k \hat{\mathbf{F}}_k^{-1} (\boldsymbol{\theta}_1 - \boldsymbol{\theta}_0)/N,$$

where

$$\hat{\mathbf{E}}_k \triangleq \frac{1}{N_k}\sum_{n=1}^{N_k} d_{k,n}^4 \left(\mathbf{Y}_n^{(k)} - \boldsymbol{\theta}_k\right)\left(\mathbf{Y}_n^{(k)} - \boldsymbol{\theta}_k\right)^T \mathbb{1}_{[0,\infty)}(d_{k,n}),$$

$$\hat{\mathbf{F}}_k \triangleq \frac{1}{N_k}\sum_{n=1}^{N_k} \frac{4d_{k,n}}{c^2}\left(\mathbf{Y}_n^{(k)} - \boldsymbol{\theta}_k\right)\left(\mathbf{Y}_n^{(k)} - \boldsymbol{\theta}_k\right)^T \mathbb{1}_{[0,\infty)}(d_{k,n}) - \mathbf{I}_{2p}\frac{1}{N_k}\sum_{n=1}^{N_k} d_{k,n}^2 \mathbb{1}_{[0,\infty)}(d_{k,n}),$$

$$\mathbf{Y}_n^{(k)} \triangleq [\text{Re}\{\mathbf{X}_n^{(k)}\}^T, \text{Im}\{\mathbf{X}_n^{(k)}\}^T]^T,$$

and $d_{k,n} \triangleq 1 - \|\mathbf{Y}_n^{(k)} - \boldsymbol{\theta}_k\|^2/c^2$. Thus, similarly to the optimization of the MT-function parameter discussed in subsection V-B, the optimal parameter $c \in C$ minimizes

$$\hat{P}_e \triangleq \pi_0 Q((t_{\text{opt}} - \kappa_0)/\sqrt{\hat{\gamma}_0}) + \pi_1 Q((\kappa_1 - t_{\text{opt}})/\sqrt{\hat{\gamma}_1}),$$

where $t_{\text{opt}}$ is the optimal threshold, which is obtained from (30) by replacing $\hat{\eta}_k^{(u)}$ and $\hat{\lambda}_k^{(u)}$ with $\kappa_k$ and $\hat{\gamma}_k$, $k=0,1$, respectively. In the simulation example the interval $C$ was set to $[0.01, 50]$.

### E. Maximum likelihood estimation of GGD parameters - Signal classification example:

In this section we provide exact implementation details of the ML estimator for the GGD parameters comprising density estimation based detector in Section VI-B. Here, similarly to Section B, we also consider the family of generalized Gaussian distributions (GGD) [s1], with densities:

$$\tilde{f}(\mathbf{X};\boldsymbol{\theta}_k) = \frac{s_k \Gamma(p)\exp\left(-\left((\mathbf{X}-\mathbf{a}_k)^H \boldsymbol{\Sigma}_k^{-1}(\mathbf{X}-\mathbf{a}_k)\right)^{s_k}\right)}{\pi^p \Gamma(p/s_k)\,|\boldsymbol{\Sigma}_k|},$$

where $k=0,1$ denotes a hypothesis index, $\Gamma(\cdot)$ is the gamma function, $\boldsymbol{\Sigma}_k \triangleq \alpha_k \mathbf{I}$, $k=0,1$, and $\boldsymbol{\theta}_k \triangleq [\alpha_k, s_k]^T$, $k=0,1$. Similarly to the signal detection example, by equating the gradient of the objective functions $J_k(\boldsymbol{\theta}_k) \triangleq \sum_{n=1}^{N_k} \log \tilde{f}(\mathbf{X}_n^{(k)};\boldsymbol{\theta}_k)$, $k=0,1$ to zero, one can verify that the ML estimators of the vectors $\boldsymbol{\theta}_k$, $k=0,1$ are the solutions of the equations

$$\hat{\alpha}_k = \left(\frac{\hat{s}_k}{pN_k}\sum_{n=1}^{N_k}\|\mathbf{X}_n^{(k)} - \mathbf{a}_k\|^{2\hat{s}_k}\right)^{1/\hat{s}_k},$$

$$\hat{s}_k = p\psi_0(p/\hat{s}_k)\left(\frac{\hat{s}_k}{N_k}\sum_{n=1}^{N_k}\frac{\|\mathbf{X}_n^{(k)} - \mathbf{a}_k\|^{2\hat{s}_k}}{\hat{\alpha}_k^{\hat{s}_k}}\log\frac{\|\mathbf{X}_n^{(k)} - \mathbf{a}_k\|^2}{\hat{\alpha}_k^{\hat{s}_k}} - 1\right)^{-1},$$

$k=0,1$, where $\psi_0(\cdot)$ is the polygamma function of order 0. The solution of these equations was obtained by fixed-point iteration. The maximum number of iterations and the stopping criterion were set to 1000 and $|J_k(\boldsymbol{\theta}_k^{(l)}) - J_k(\boldsymbol{\theta}_k^{(l-1)})|/|J_k(\boldsymbol{\theta}_k^{(l-1)})| < 10^{-6}$, respectively, where $\boldsymbol{\theta}_k^{(l)}$ denotes the estimates of $\boldsymbol{\theta}_k$ at iteration index $l$. Both the initial conditions of $\hat{s}_0$ and $\hat{s}_1$ were set to 1.

### F. Simulations with small sample size:

In this section, we repeat the numerical examples in Subsections VI-A and VI-B for small sample size of $N=20$. We note that here, ROC curve analysis was performed for SNR $=-5$ [dB] for both Gaussian and non-Gaussian noise. Furthermore, the threshold values of all compared tests were determined via $(10^5)$ Monte-Carlo simulations. Observing Figs. S1 and S2, one sees that for both signal detection and classification examples, the proposed MT-GQLRT$_{\text{opt}}$ attains the best detection and classification performance, except the omniscient LRT, which assumes complete knowledge of the likelihood function under each hypothesis (Recall that MT-GQLRT$_{\text{opt}}$ and MT-GQLRT$_{\text{sub}}$ are the optimal and suboptimal MT-GQLRT implementations, respectively, discussed in Subsection VI-A and VI-B). Moreover, one can notice that under the signal detection example, the MT-GQLRT$_{\text{sub}}$ performs similarly to the NSDD-GQLRT for Gaussian noise (Figs. 1(a) and 1(b)). For the non-Gaussian noise (Figs. 1(c) and 1(d)), it outperforms all compared tests, excluding the MT-GQLRT$_{\text{opt}}$ and the omniscient LRT. Under the signal classification example (Fig. S2), the MT-GQLRT$_{\text{sub}}$ outperforms all robust GQLRT alternatives that do not use training sequences (except for threshold determination).

### G. Signal detection example: A modified scale-invariant version of the MT-GQLRT:

In this section, a modified scale invariant version of the proposed test (37) is developed. In order to induce scale-invariance, the test-statistic $T_u'$ (36) is normalized by an observation-dependent factor. This factor is identically distributed over the considered hypotheses, and therefore, it converges almost surely to the same constant value under $H_0$ and $H_1$ when the sample size approaches infinity. More specifically, we define the following normalized test-statistic:

$$T_u'' \triangleq \frac{T_u'}{D} = \frac{\mathbf{a}^H \hat{\mathbf{C}}^{(u)} \mathbf{a}}{D}, \tag{S-9}$$

where $D \triangleq \text{trace}\{\mathbf{P}_{\mathbf{a}}^{\perp} \hat{\mathbf{C}}^{(u)}\}$, and $\hat{\mathbf{C}}^{(u)}$ is defined below (36). The decision rule based on this modified test-statistic is given by:

$$T_u'' \underset{H_0}{\overset{H_1}{\gtrless}} t. \tag{S-10}$$

Interestingly, the test (S-10) is a measure-transformed version of the Gauss-Gauss detector [s4] (Eq. 14) for the special case of spatially white interference.

In the following we show that the detector (37) and its modified version (S-10) have the same asymptotic power at a

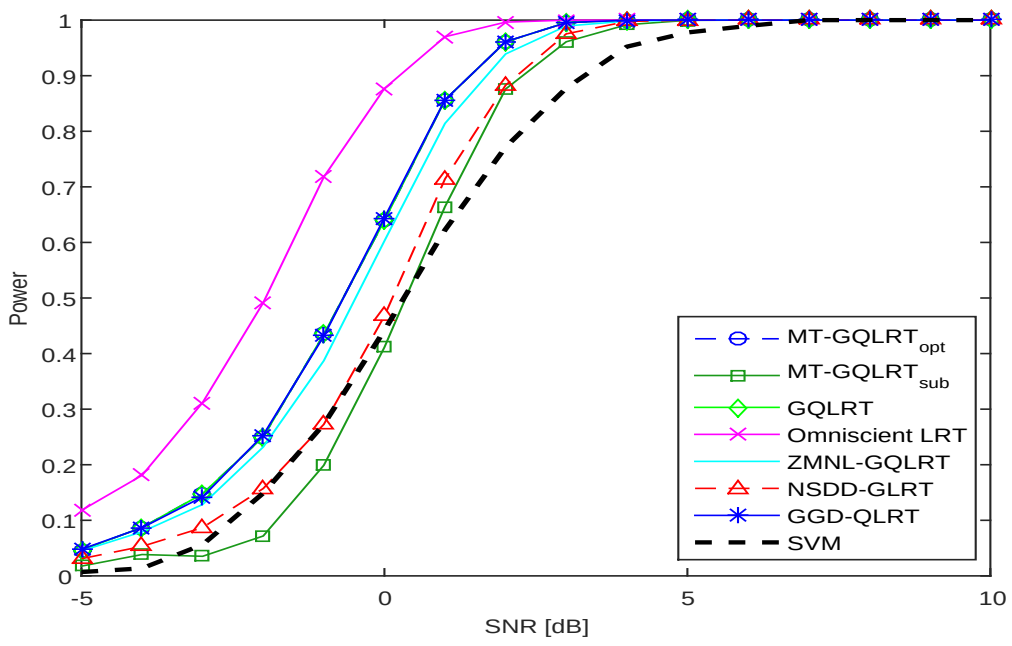

(a)

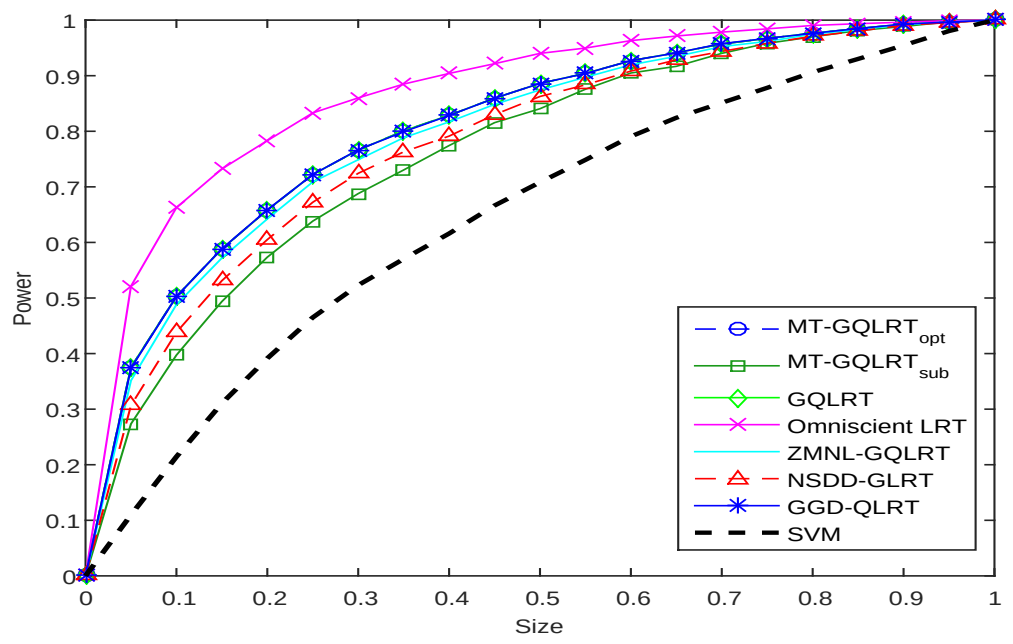

(b)

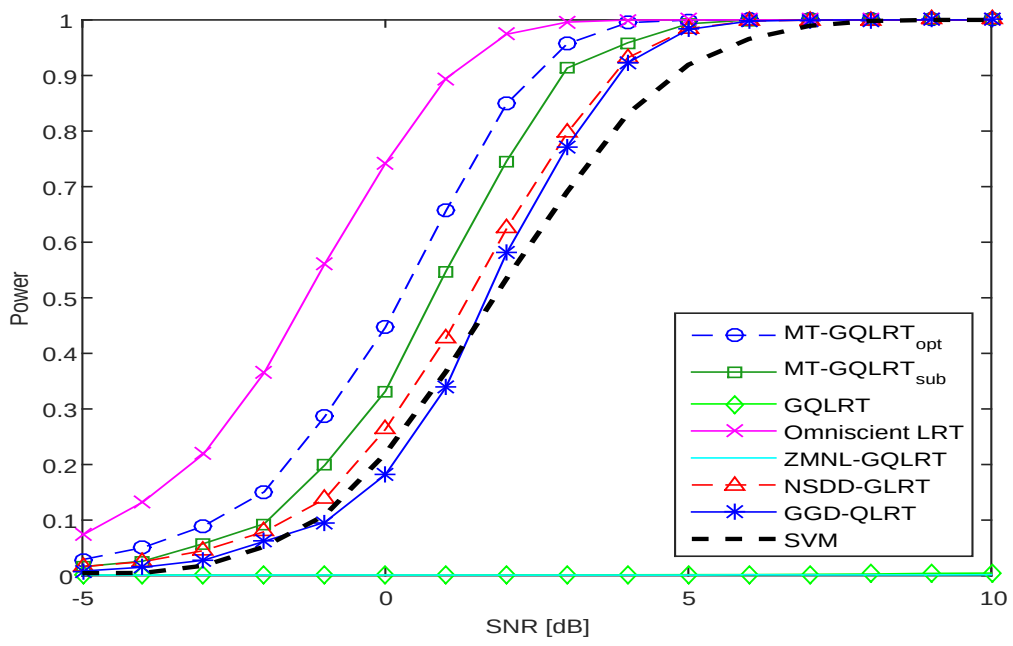

(c)

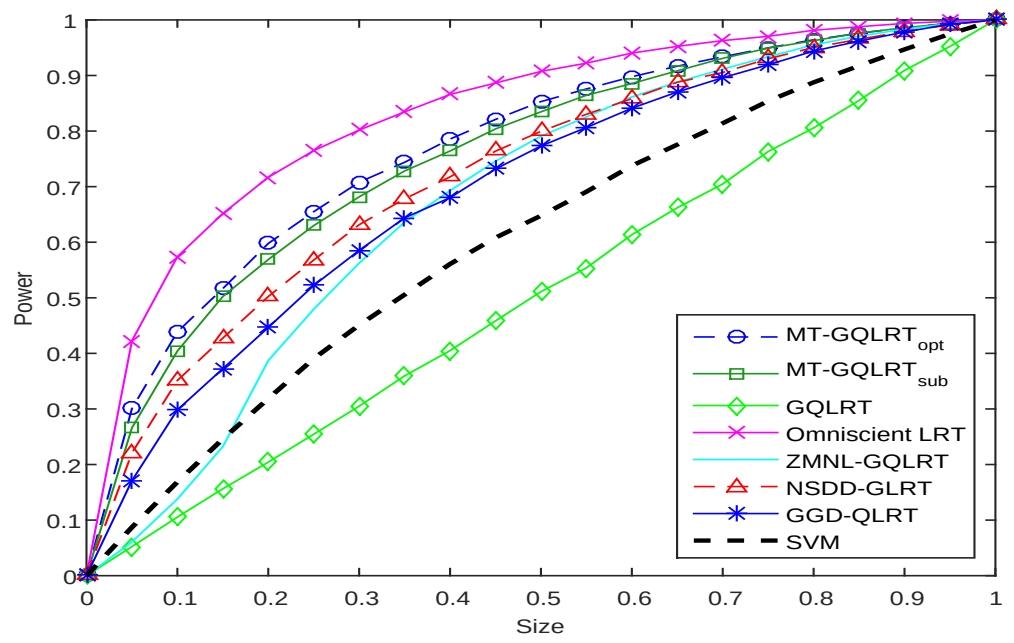

(d)

Fig. S1. **Signal detection with small sample size:** (a) + (b) in Gaussian noise, (c) + (d) in non-Gaussian noise.

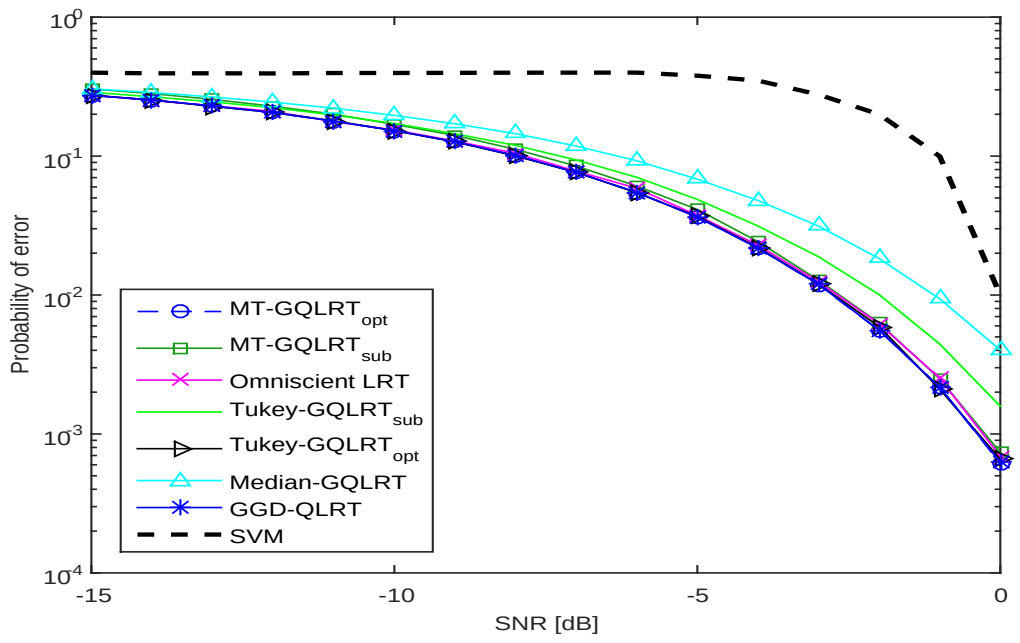

(a)

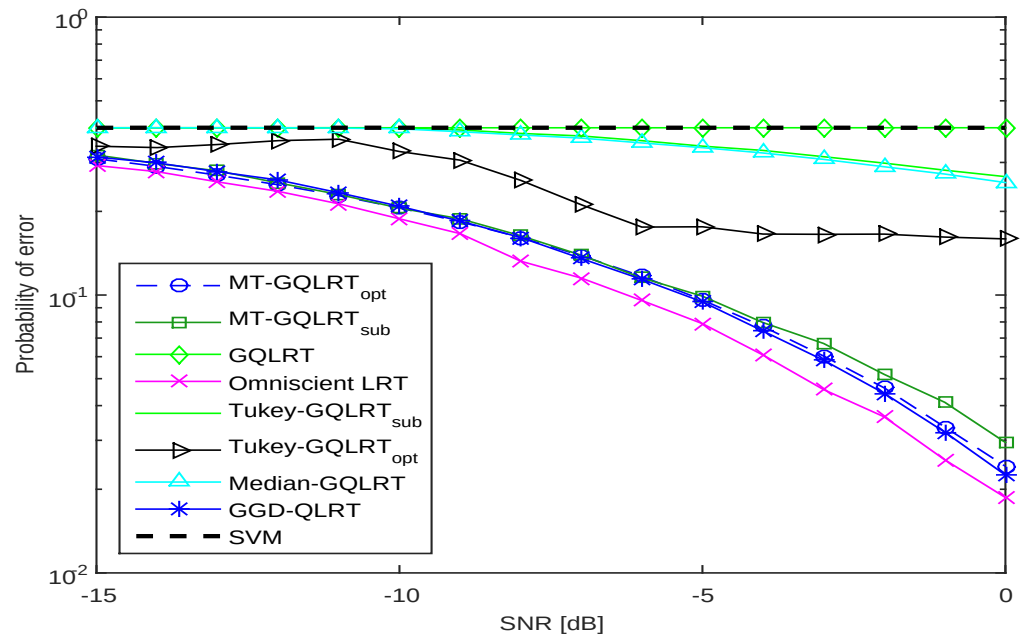

(b)

Fig. S2. **Signal classification with small sample size:** (a) in Gaussian noise, (b) in non-Gaussian noise.

fixed false alarm rate. Under the parameterized Gaussian MT-function (42), this property justifies the selection of the width parameter $\omega$ via the selection rule (65). Using (34), (35), the last equality in (S-7), and the consistency of the empirical MT-mean and MT-covariance (see Proposition 2 in [s5]), one can verify that

$$D \xrightarrow{\text{w.p.1}} (p-1)\, r_1^{(u)} \triangleq \gamma \quad \text{as} \quad N \rightarrow \infty$$

under both $H_0$ and $H_1$, where $r_1^{(u)}$ is defined below (S-7). Therefore, by (36), (S-9), Theorem 1 (which implies asymptotic normality of $T'_u$), and Slutsky's theorem [s6], we conclude that under $H_k$, $k \in \{0,1\}$:

$$\frac{T''_u - \bar{\eta}_k^{(u)}}{\sqrt{\bar{\lambda}_k^{(u)}}} = A_1 \frac{T'_u - \tilde{\eta}_k^{(u)}}{\sqrt{\tilde{\lambda}_k^{(u)}}} + A_2 \xrightarrow[N \rightarrow \infty]{D} \mathcal{N}\left(0,1\right),$$

where $\bar{\eta}_k^{(u)} \triangleq \tilde{\eta}_k^{(u)}/\gamma$, $\bar{\lambda}_k^{(u)} \triangleq \tilde{\lambda}_k^{(u)}/\gamma^2$, $\tilde{\eta}_k^{(u)} \triangleq \frac{\eta_k^{(u)} - c_1^{(u)}}{c_2^{(u)}}$, $\tilde{\lambda}_k^{(u)} \triangleq \frac{\lambda_k^{(u)}}{(c_2^{(u)})^2}$, the constants $c_1^{(u)}$ and $c_2^{(u)}$ are defined above (36), and $\eta_k^{(u)}$ and $\lambda_k^{(u)}$ are defined in (21) and (22), respectively. The random variables $A_1 \triangleq \gamma/D$ and $A_2 \triangleq \tilde{\eta}_k^{(u)}(\gamma/D - 1)/\sqrt{\bar{\lambda}_k^{(u)}}$. Hence, the asymptotic power of the modified test (S-10) at a fixed asymptotic size $\alpha$ is given by:

$$\bar{\beta}_u^{(\alpha)} = Q\left(\frac{\bar{\eta}_0^{(u)} - \bar{\eta}_1^{(u)} + \sqrt{\bar{\lambda}_0^{(u)}}Q^{-1}(\alpha)}{\sqrt{\bar{\lambda}_1^{(u)}}}\right) \quad \text{(S-11)}$$

$$= Q\left(\frac{\eta_0^{(u)} - \eta_1^{(u)} + \sqrt{\lambda_0^{(u)}}Q^{-1}(\alpha)}{\sqrt{\lambda_1^{(u)}}}\right) = \beta_u^{(\alpha)},$$

which by corollary 1 and (36) is exactly the asymptotic power of (37) under the same test size.

In the following Proposition, we show that the test-statistic (S-9) implemented with the Gaussian MT-function $u_G(\mathbf{x};\omega)$ (42) is scale invariant when the width parameter $\omega$ is selected according to the selection rule (65). This result guarantees constant false alarm rate (CFAR) property w.r.t. the noise power $\sigma^2_{\mathbf{W}}$.

**Proposition 1.** *Let $\omega^2_{\mathbf{Y}} = \left(1/\sqrt{\zeta} - 1\right)\hat{\sigma}^2_{\mathbf{Y}}$ where $\zeta$ and $\hat{\sigma}^2_{\mathbf{Y}}$ are defined below (65). Then, for the Gaussian MT-function (42) and any constant $s \in \mathbb{C}$, the modified test-statistic (S-9) satisfies:*

$$T''_{u_G}\left(\{\mathbf{X}_n\}_{n=1}^N;\omega^2_{\mathbf{Y}}\right) = T''_{u_G}\left(\{s\mathbf{X}_n\}_{n=1}^N;\omega^2_{s\mathbf{Y}}\right), \quad \text{(S-12)}$$

*where $\omega^2_{s\mathbf{Y}} = \left(1/\sqrt{\zeta} - 1\right)\hat{\sigma}^2_{s\mathbf{Y}}$.*

*Proof.* Since the MAD estimator of variance is linear w.r.t. scale, we have that $\omega^2_{s\mathbf{Y}} = |s|^2\omega^2_{\mathbf{Y}}$. Hence, by (12), (13) and (42) we conclude that the empirical measure-transformed autocorrelation, defined below (36), that comprise the test statistic (S-9), satisfies:

$$\hat{\mathbf{C}}^{(u_G)}\left(\{s\mathbf{X}_n\}_{n=1}^N;\omega^2_{s\mathbf{Y}}\right) = |s|^2\frac{\sum_{n=1}^N u_G\left(s\mathbf{X}_n;\omega^2_{s\mathbf{Y}}\right)\mathbf{X}_n\mathbf{X}_n^H}{\sum_{n=1}^N u_G\left(s\mathbf{X}_n;\omega^2_{s\mathbf{Y}}\right)}$$
$$= |s|^2\frac{\sum_{n=1}^N u_G\left(\mathbf{X}_n;\omega^2_{\mathbf{Y}}\right)\mathbf{X}_n\mathbf{X}_n^H}{\sum_{n=1}^N u_G\left(\mathbf{X}_n;\omega^2_{\mathbf{Y}}\right)}$$
$$= |s|^2\hat{\mathbf{C}}^{(u_G)}\left(\{\mathbf{X}_n\}_{n=1}^N;\omega^2_{\mathbf{Y}}\right). \quad \text{(S-13)}$$

Finally, by (S-9) and (S-13)

$$T''_{u_G}\left(\{s\mathbf{X}_n\}_{n=1}^N;\omega^2_{s\mathbf{Y}}\right) = \frac{\mathbf{a}^H\hat{\mathbf{C}}^{(u)}\left(\{s\mathbf{X}_n\}_{n=1}^N;\omega^2_{s\mathbf{Y}}\right)\mathbf{a}}{\text{trace}\{\mathbf{P}^{\perp}_{\mathbf{a}}\hat{\mathbf{C}}^{(u)}\left(\{s\mathbf{X}_n\}_{n=1}^N;\omega^2_{s\mathbf{Y}}\right)\}}$$
$$= \frac{\mathbf{a}^H\hat{\mathbf{C}}^{(u)}\left(\{\mathbf{X}_n\}_{n=1}^N;\omega^2_{\mathbf{Y}}\right)\mathbf{a}}{\text{trace}\{\mathbf{P}^{\perp}_{\mathbf{a}}\hat{\mathbf{C}}^{(u)}\left(\{\mathbf{X}_n\}_{n=1}^N;\omega^2_{\mathbf{Y}}\right)\}}$$
$$= T''_{u_G}\left(\{\mathbf{X}_n\}_{n=1}^N;\omega^2_{\mathbf{Y}}\right).$$

□

The modified test (S-10) will be called here "MT-GQLRT$_{\text{mod}}$". In the following, we examine its detection performance. To do so, we repeated the second simulation example in Subsection VI-A, where the MT-GQLRT$_{\text{mod}}$ replaces the suboptimal MT-GQLRT implementation, called MT-GQLRT$_{\text{sub}}$, under which the same selection rule (65) is applied. Similarly to the MT-GQLRT$_{\text{sub}}$, the threshold is determined via Monte-Carlo simulations. Here, the relative local power sensitivity in (65) was set to to $r = 0.99$. All the other settings are remained as described in subsection VI-A. Observing Figs. S3(a)-S3(c) and S4(a)-S4(c), one sees that the modified test (S-10) outperforms the other compared tests that do not use training sequences for parameter tuning (other than the threshold).

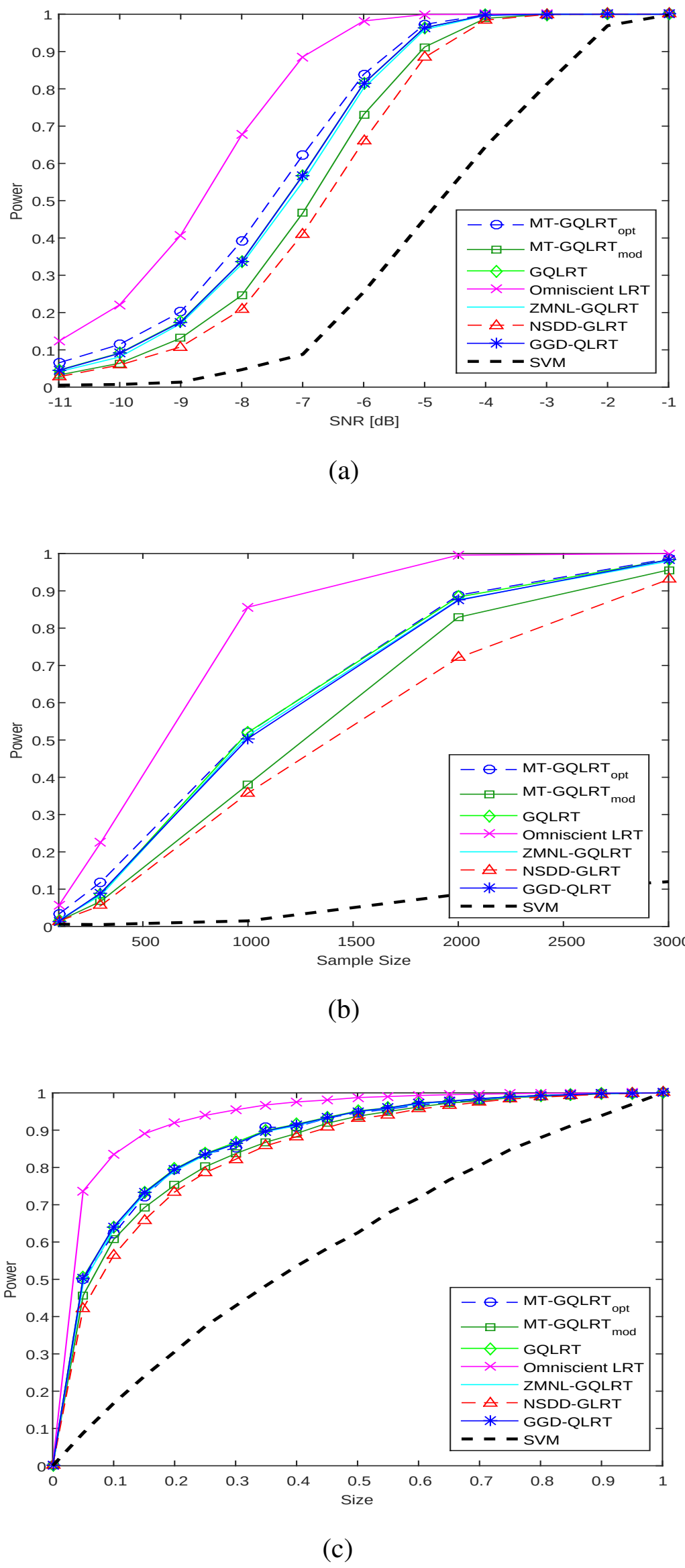

Fig. S3. Signal detection in Gaussian noise using the modified test (S-10)

Next, to illustrate CFAR property w.r.t noise power, empirical test sizes were obtained for modified values of the noise power $\sigma^2_{\mathbf{W}}$ when the threshold in (S-10) was determined for $\sigma^2_{\mathbf{W}} = 1$. Observing Fig. S5, one sees that indeed, the proposed detector has a CFAR property w.r.t. noise power.

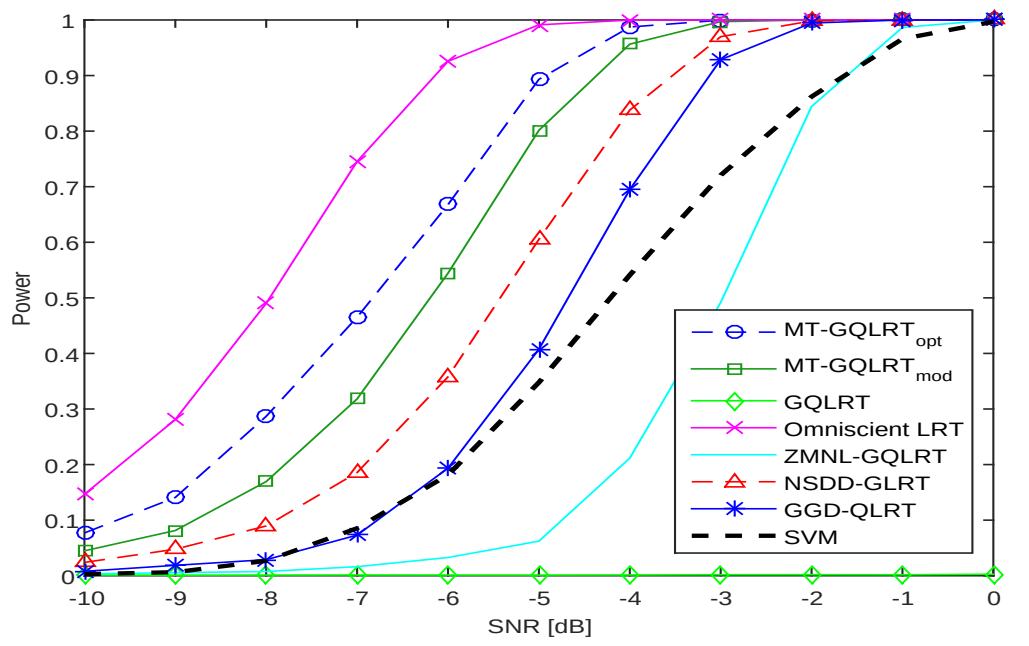

(a)

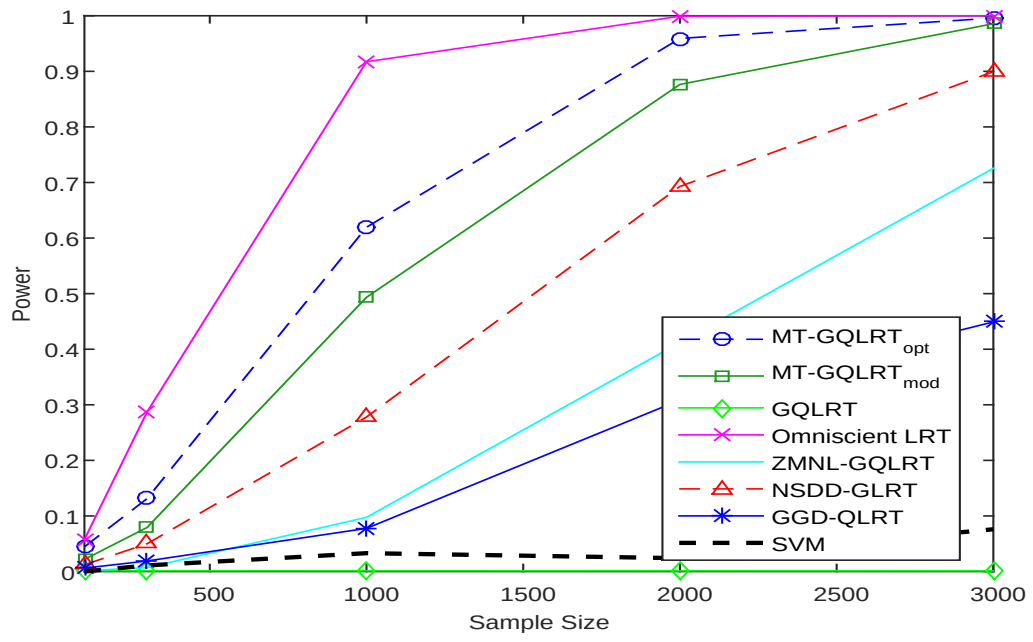

(b)

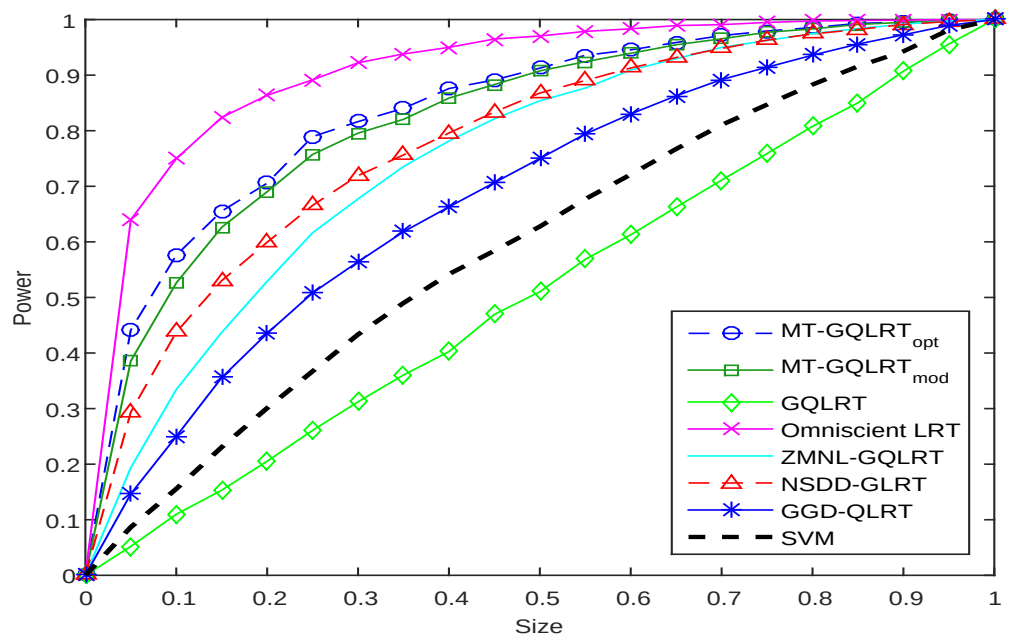

(c)

Fig. S4. Signal detection in non-Gaussian noise using the modified test (S-10)

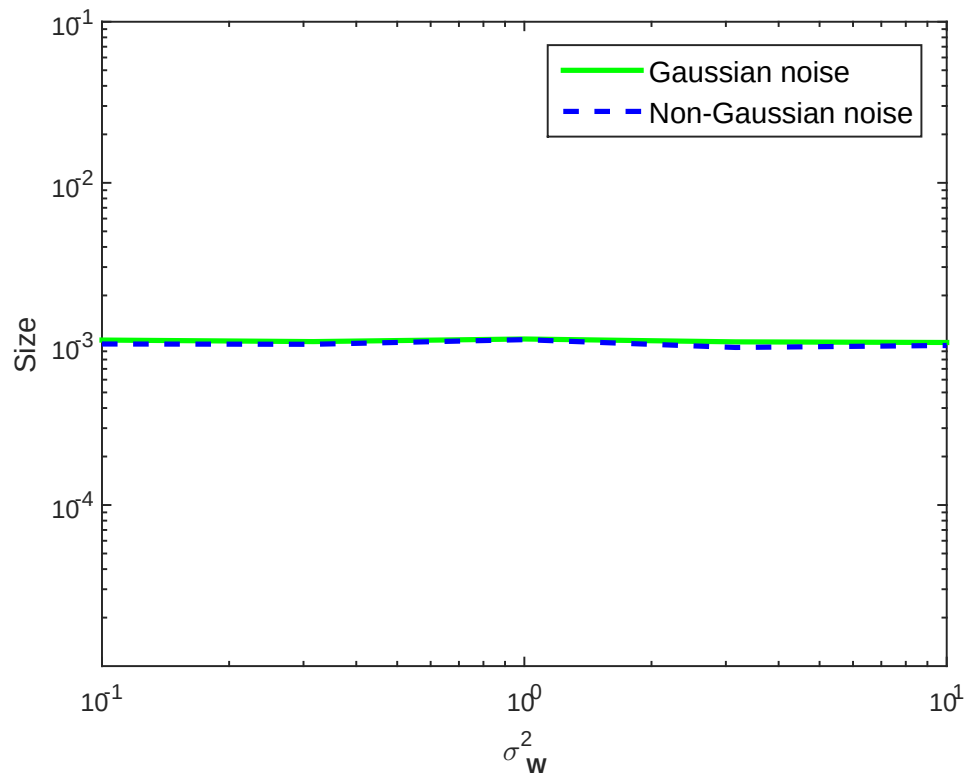

Fig. S5. Empirical CFAR analysis of the test (S-10)